\documentclass[%
 reprint,
superscriptaddress,
 amsmath,amssymb,
 aps,
prb,
]{revtex4-1}

\usepackage{graphicx}
\usepackage{subcaption}
\usepackage{dcolumn}
\usepackage{bm}
\usepackage{xcolor}
\usepackage{wasysym}

\usepackage{multirow}

\begin{document}


\title{Effects of uniaxial pressure on the spin ice Ho\textsubscript{2}Ti\textsubscript{2}O\textsubscript{7} }

\author{R. Edberg}
\affiliation{Physics Department, KTH Royal Institute of Technology, Sweden}

\author{I. M. B. Bakke}
\affiliation{Centre for Materials Science and Nanotechnology, Department of Chemistry, University of Oslo, Norway}

\author{H. Kondo}
\affiliation{Faculty of Engineering, Kyushu Institute of Technology, Kitakyushu, Fukuoka 804-8550, Japan}

\author{L. \O rduk Sandberg}
\affiliation{Niels Bohr Institute, University of Copenhagen, Denmark}

\author{M. L. Haubro}
\affiliation{Niels Bohr Institute, University of Copenhagen, Denmark}
 
\author{M. Guthrie}%
 \affiliation{European Spallation Source ERIC, 22363 Lund, Sweden}
\affiliation{University of Edinburgh, School of Physics and Astronomy and Centre for Science at Extreme Conditions}

\author{A. T. Holmes}%
 \affiliation{European Spallation Source ERIC, 22363 Lund, Sweden}

\author{J. Engqvist}%
\affiliation{Division of Solid Mechanics, Lund University, P.O. Box 118, SE-221 00 Lund, Sweden}

\author{A. Wildes}
\affiliation{Institut Laue-Langevin, 38042 Grenoble, France}

\author{K. Matsuhira}%
\affiliation{Faculty of Engineering, Kyushu Institute of Technology, Kitakyushu, Fukuoka 804-8550, Japan}

\author{K. Lefmann}%
\affiliation{Niels Bohr Institute, University of Copenhagen, Denmark}

\author{P. P. Deen}%
\affiliation{Niels Bohr Institute, University of Copenhagen, Denmark}
\affiliation{European Spallation Source ERIC, 22363 Lund, Sweden}

\author{M. Mito}%
\affiliation{Faculty of Engineering, Kyushu Institute of Technology, Kitakyushu, Fukuoka 804-8550, Japan}

\author{P. Henelius}%
\affiliation{Physics Department, KTH Royal Institute of Technology, Sweden}
 \affiliation{Faculty of Science and Engineering,  \r{A}bo Akademi University, \r{A}bo, Finland}

\date{\today}

\begin{abstract}

The spin ice materials Ho\textsubscript{2}Ti\textsubscript{2}O\textsubscript{7} and Dy\textsubscript{2}Ti\textsubscript{2}O\textsubscript{7} are experimental and theoretical exemplars of highly frustrated magnetic materials. However, the effects of an applied uniaxial pressure are not well studied, and here we report magnetization measurements of Ho\textsubscript{2}Ti\textsubscript{2}O\textsubscript{7} under uniaxial pressure applied in the $[001]$, $[111]$ and $[110]$ crystalline directions. The basic features are captured by an extension of the dipolar spin ice model.  We find a good match between our model and measurements with pressures applied along two of the three directions, and extend the framework to discuss the influence of crystal misalignment for the third direction. The parameters determined from the magnetization measurements reproduce neutron scattering measurements we perform under uniaxial pressure applied along the $[110]$ crystalline direction. In the detailed analysis we include the recently verified susceptibility dependence of the demagnetizing factor. Our work demonstrates the application of a moderate applied pressure to modify the magnetic interaction parameters. The knowledge can be used to predict critical pressures needed to induce new phases and transitions in frustrated materials, and in the case of Ho\textsubscript{2}Ti\textsubscript{2}O\textsubscript{7} we expect a transition to a ferromagnetic ground state for uniaxial pressures above $3.3\textup{ GPa}$.

\end{abstract}

\pacs{Valid PACS appear here}
\maketitle


\section{Introduction}

Higly frustrated magnets display a rich variety of exotic groundstates and excitations\cite{Springer_frust_book, Castelnovo_Nature, Brooks, Henley_Coulomb}. A primary reason for this diversity is that the physical properties of the system are frequently determined by a delicate balance of weaker interactions of similar strength. Following Anderson's original classification\cite{anderson87} there are two major classes of frustrated systems; one in which the lattice plays a dominant role in frustrating the system, and another where frustration arises due to competing interactions. Recently it was realized that even in the first case, competing interactions may unexpectedly refrustrate the system, as was found for the spin ice materials Ho\textsubscript{2}Ti\textsubscript{2}O\textsubscript{7} (HTO)  and Dy\textsubscript{2}Ti\textsubscript{2}O\textsubscript{7} (DTO)\cite{henelius16}. Therefore, it is of interest to find physical realizations of a wide variety of Hamiltonians, since even small alterations of the interactions may  yield novel physical properties and phases.

In order to explore the vast parameter space of frustration there is a constant drive to synthesize new and promising frustrated materials\cite{Springer_frust_book,hallas15, biesner20}, each representing a unique set of interaction coefficients determined by, for  example, crystal fields, ionic magnetic moments, inter-ion distances and  atomic overlaps. A different approach to probing frustration is to alter the interactions of a given material. One way to do so is the application of external pressure, which alters the  position and thereby also the dipolar interactions and atomic overlaps of the ions of the material. While there are many observations of novel states and phenomena induced by pressure\cite{mirebeau02,kim17,zyvagin19}, progress is hampered by a number of experimental challenges that are exacerbated when studying phenomena at cryogenic temperatures. On the theoretical side there are open questions, since it is not easy  to predict the effects of external pressure on, for example, exchange interaction parameters. 

In the present study, we apply uniaxial external pressure to increase the parameter space we can explore using the parent material HTO. The physical properties of the multiaxial Ising materials HTO and DTO  show a strong directional dependence on, for example an applied magnetic field~\cite{sondhi04,fennell05}. Therefore we choose to use uniaxial pressure rather than isotropic hydrostatic pressure in this study. Our starting point are relatively straight forward  magnetization measurements to  determine the evolution of the exchange parameters under pressure and then we validate our results by comparing the derived model to neutron scattering experiments. In this study, we demonstrate the feasibility and limitations of describing the measurements  of HTO under pressure using  an effective model with a single pressure dependent parameter. While our application of relatively low pressure did not induce any changes in the order of the material, it shows the feasibility of exploratory experimental studies using magnetization measurements. Such measurements can be used to determine the  evolution the model parameters and enable theoretical predictions of pressures necessary to alter the delicate balance of interactions in frustrated materials enough to cross phase boundaries and induce new phases.

Over the last 15 years HTO and DTO have become model compounds for studying classically frustrated systems featuring residual ground state entropy\cite{Ramirez}, topological magnetic monopole excitations\cite{Castelnovo_Nature}, and slow low-temperature dynamics\cite{jaubert09, Revell13}. The theoretical description in terms of the dipolar spin-ice model (DSM)\cite{denHertog, Siddharthan,Yavo08} captures the experimental features at a quantitative level and  DTO and HTO are now some of the best characterized frustrated materials. This makes them an ideal starting point for a systematic exploration of the evolution of frustrated systems under pressure. In an earlier study, measurements of the pressure induced changes in the magnetization of DTO were reported\cite{MITO}. These results were first  modeled theoretically using a nearest-neighbor exchange interaction model\cite{Jaubert1}. Recently, the effects of the dipolar interactions were included in the theory, and the measurements were modeled  using a single parameter in the  DSM\cite{OurArticle}. Since DTO has a large neutron absorption cross section, neutron scattering measurements were performed on HTO, rather than DTO, but the observed changes in scattering intensity were modeled within the same framework.

One shortcoming of the previous study\cite{OurArticle} was that magnetization measurements of HTO under pressure were missing, and it was not possible to connect the parameters determined from the evolution of the magnetization, measured for DTO, with the changes in the scattering intensity, determined for HTO\cite{OurArticle}. In this study, we remedy the situation and report magnetization measurements on HTO, which we can use to independently determine the pressure dependent parameter in the theory for HTO. Our main result is that the independent analysis of both the magnetization and the neutron scattering measurements lead to the very similar values for the pressure dependent parameter. We therefore verify the feasibility of our approach to use more straight forward magnetization measurements to determine the explicit evolution of  model parameters to predict possible phase transitions before performing the more demanding neutron scattering experiments under pressure.

\section{Experimental Method and Results}
\subsection{Magnetization measurements}\label{section:MagnetizationExperiment}
High-quality samples are essential in high-pressure studies, and we synthesized HTO single crystals using the floating zone technique at the image furnace at Lund University. The crystals were grown from the bottom up with a growth rate of $3\textup{ mm/hr}$. Visual inspection of the as-grown crystals revealed no color change along the growth direction as seen by others.\cite{Ghasemi}

\begin{figure}[h!]
	\begin{subfigure}[t]{0.5\textwidth}
		\centering
		\includegraphics[width=0.85\linewidth]{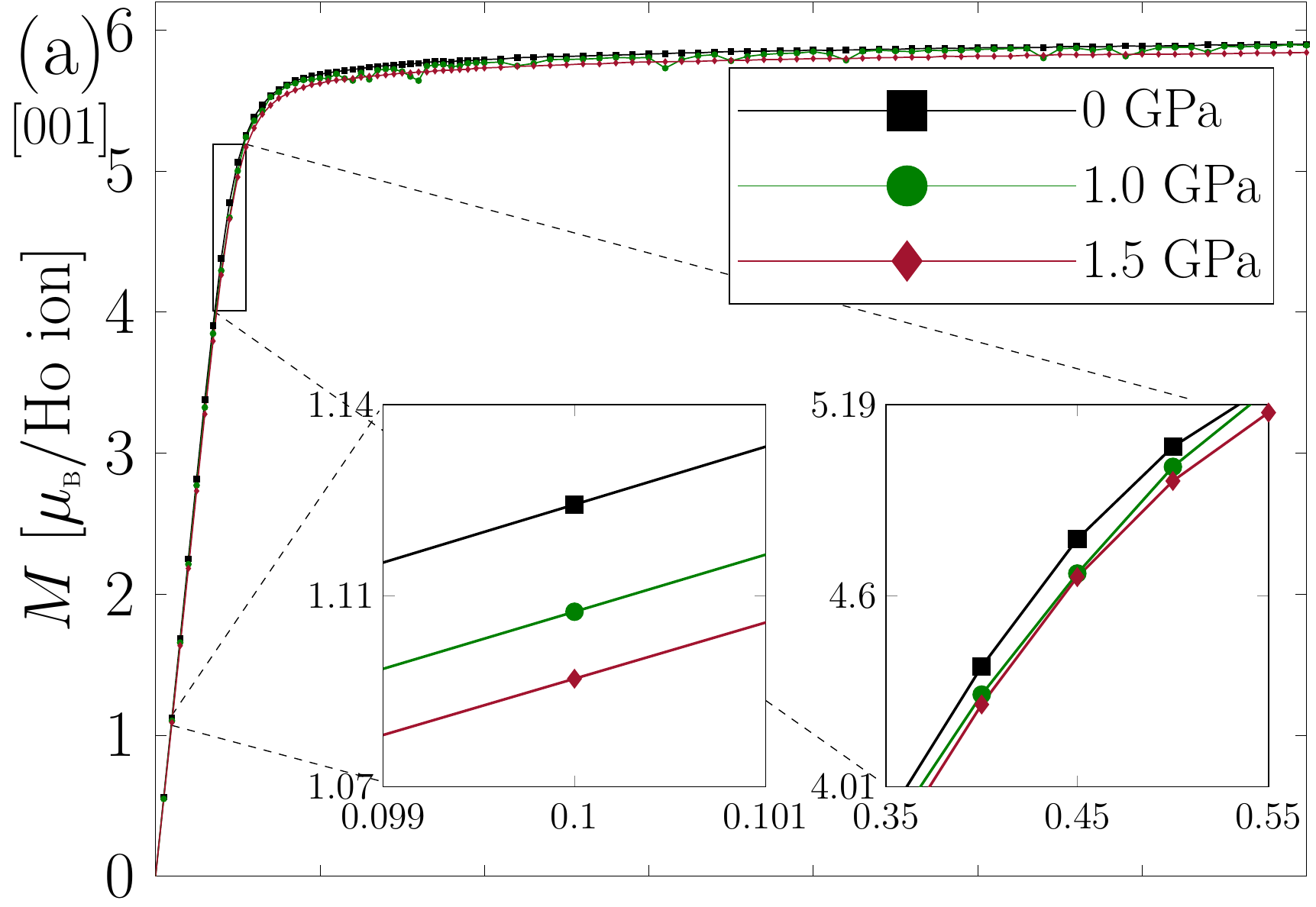}
	\end{subfigure}
	\begin{subfigure}[t]{0.5\textwidth}
		\centering
		\includegraphics[width=0.85\linewidth]{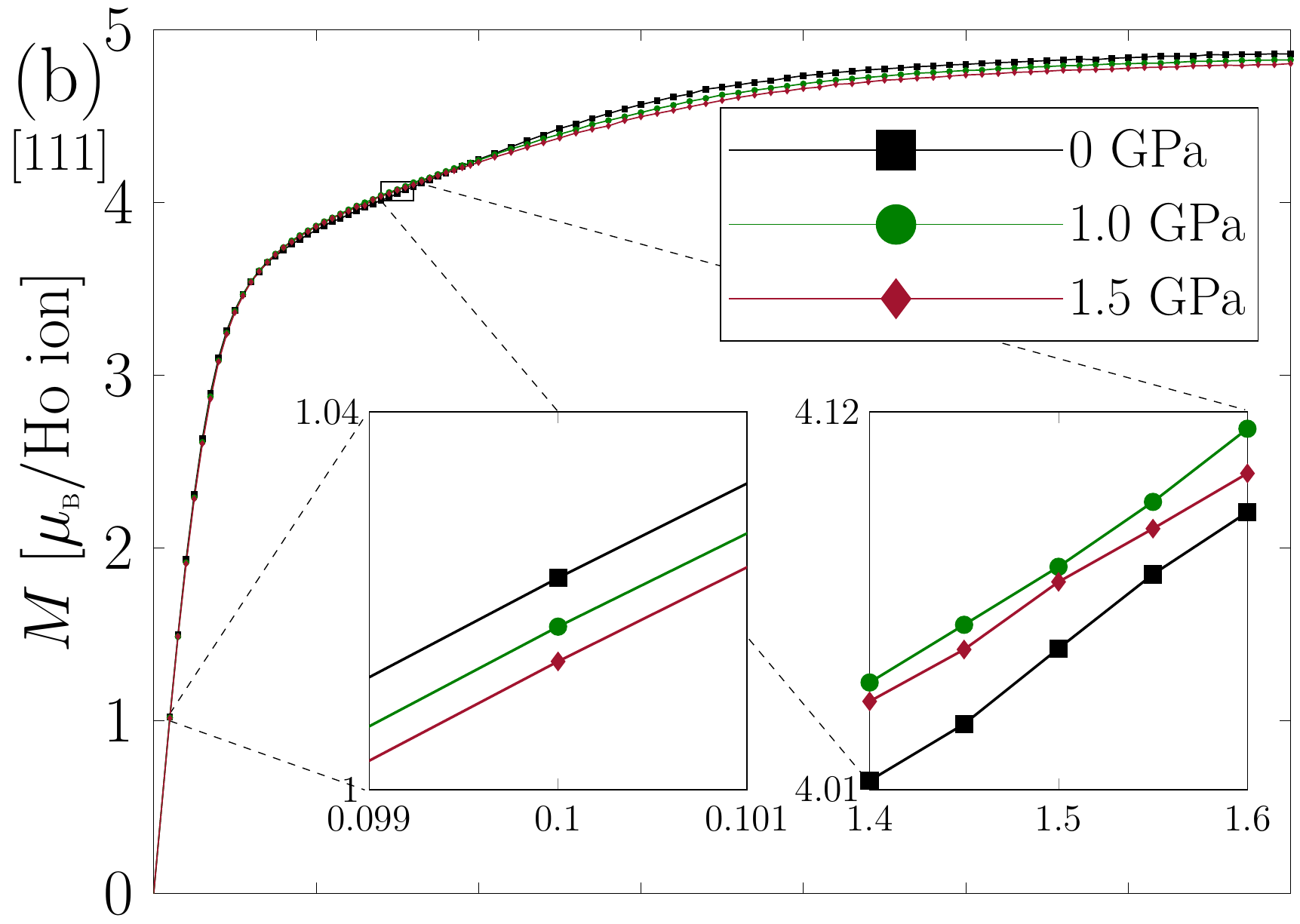}
	\end{subfigure}
	\begin{subfigure}[t]{0.5\textwidth}
		\centering
		\hspace{0.04mm}
		\includegraphics[width=0.858\linewidth]{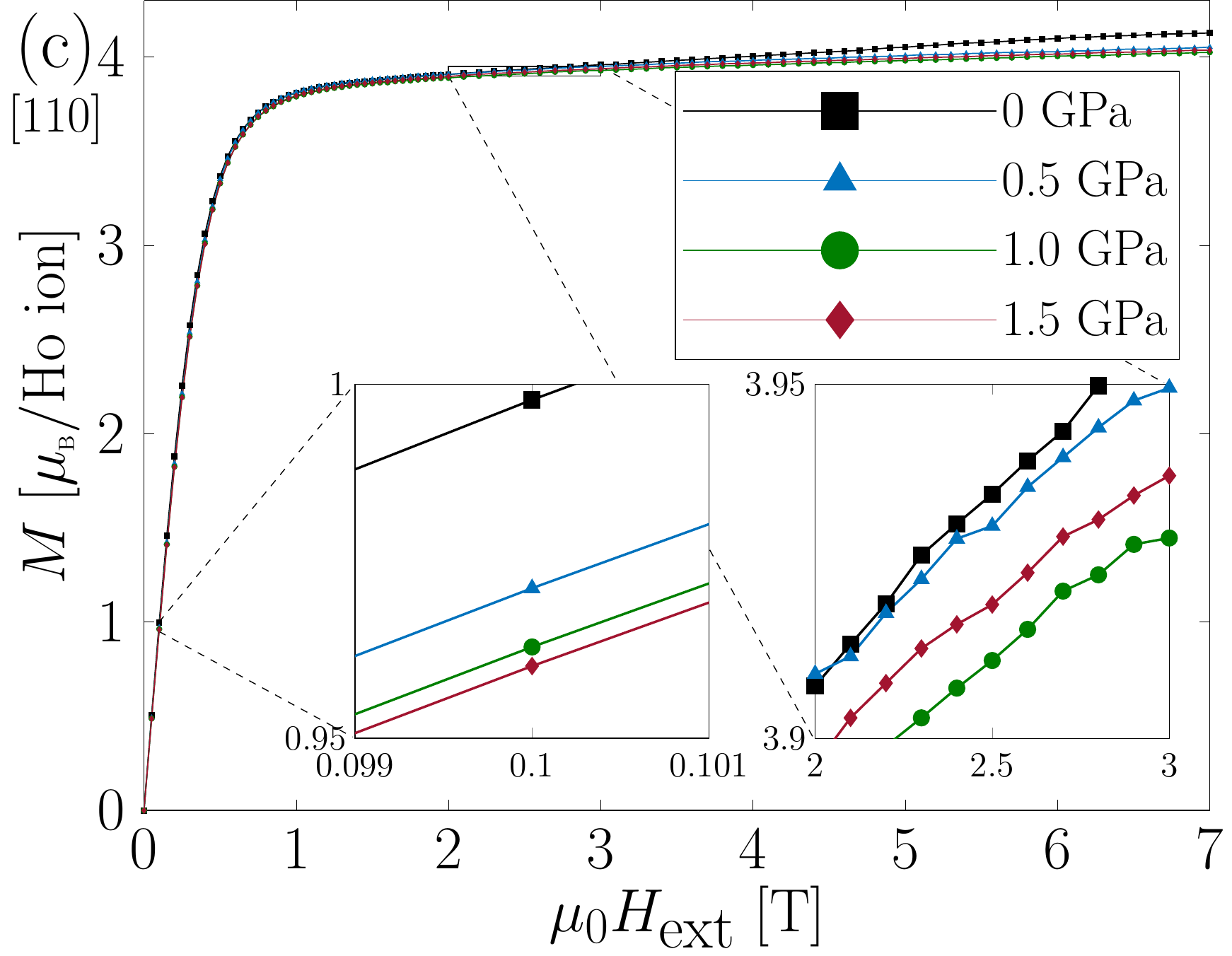}
	\end{subfigure}
	\caption{Sample magnetic moment per holmium ion as function of external magnetic field $H_\textup{ext}$, without demagnetization corrections. Pressure and magnetic field along the (a) $[001]$, (b) $[111]$ and (c) $[110]$ crystalline direction. The different curves indicate different pressure applied to the sample, all at sample temperature $1.83\textup{ K}$. The insets show the changes in magnetic moment at small and moderate field.}
	\label{fig:rawData111.001}
\end{figure}

We ascertained the high quality of the single crystals using x-ray and neutron diffraction \cite{D9_data} and will publish the  details of the growth and resultant crystalline state elsewhere. To ensure an optimum shape for the application of uniaxial pressure, the single crystals were cut into cylinders of diameter $2.0\pm0.05 \textup{ mm}$ and height $3.0\pm0.05\textup{ mm}$. The crystals were cut with the major axis of the cylinders along the crystalline $[001]$, $[110]$ and $[111]$ directions. The quality after cutting was asserted using x-ray diffraction. The crystalline direction of the major axis was ascertained after cutting the crystals. For each crystal, we applied uniaxial pressure and magnetic field along the cylinder major axis. We measured the magnetic moment at $p_{001}=p_{111}=0,\;1.0\pm0.2$ and $1.5\pm0.3\textup{ GPa}$ for the $[001]$ and $[111]$ crystalline direction and at $p_{110}=0,\;0.5\pm0.1,\;1.0\pm0.2$ and $1.5\pm0.3$ $\textup{ GPa}$ for the $[110]$ crystalline direction. We also measured the DC susceptibility at pressures $p_{001}=p_{111}=0,\;0.5\pm0.1,\;1.0\pm0.2$ and $1.3\pm0.2\textup{ GPa}$ using a probing field of $0.01\textup{ T}$ along the cylinder major axis.

Using an epoxy resin (Stycast 1266, Ablestick Japan Co., Ltd.) we  placed the crystals into a piston- cylinder type of pressure cell (CR-PSC-KY05-1, Kyowa- Seisakusho Co., Ltd.), which can be inserted into a superconducting quantum interference device (SQUID) magnetometer (Quantum Design MPMS-XL). The combination of Stycast and the above uniaxial pressure enables the Poisson effect to be excluded and thus the shrinkage ratio in the direction perpendicular to the load can be ignored\cite{PaperPascaleWantedToCiteForZeroPoissonRatio_Vol6921642000}. The quoted value of the pressure is at liquid helium temperature, after considering the thermal shrinkage. The pressure value at liquid helium temperature was estimated from the shift of the superconducting transition temperature of lead under applied pressure\cite{MITO_superconductingLead}. Due to the symmetric design of the upper and lower part of the sample chamber the magnetic background was negligible for the large magnetic signal of HTO. The measurements at different pressure were performed on the same respective crystal for each direction and a maximum field of $H_\textup{ext}=7\textup{ T}$ was used. 


 \begin{figure}[h!] 
	\begin{subfigure}[t]{0.5\textwidth}
		\centering
		\includegraphics[width=0.8335\linewidth]{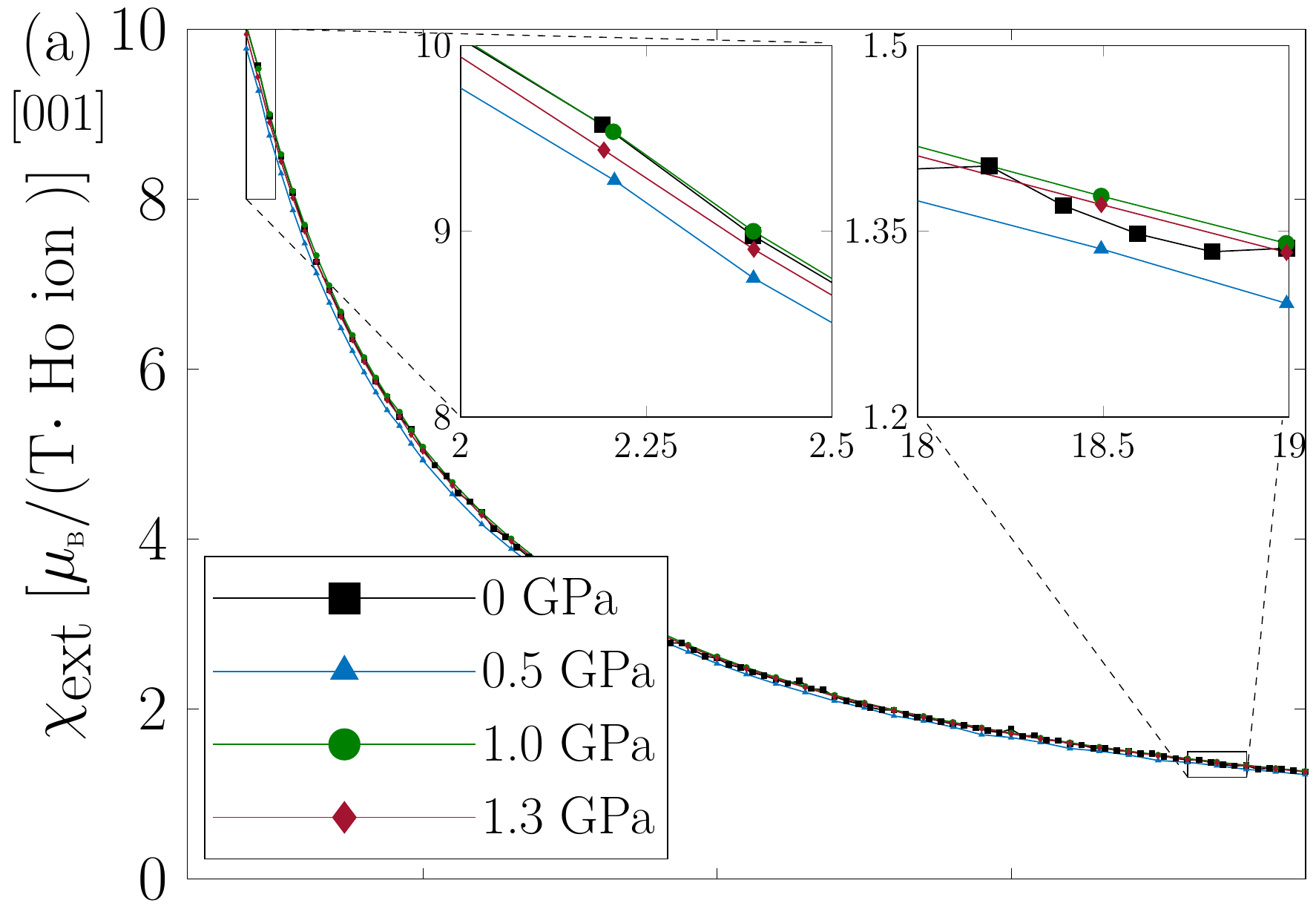}
	\end{subfigure}
	\begin{subfigure}[t]{0.5\textwidth}
		\centering
		\hspace{0.3mm}
		\includegraphics[width=0.85\linewidth]{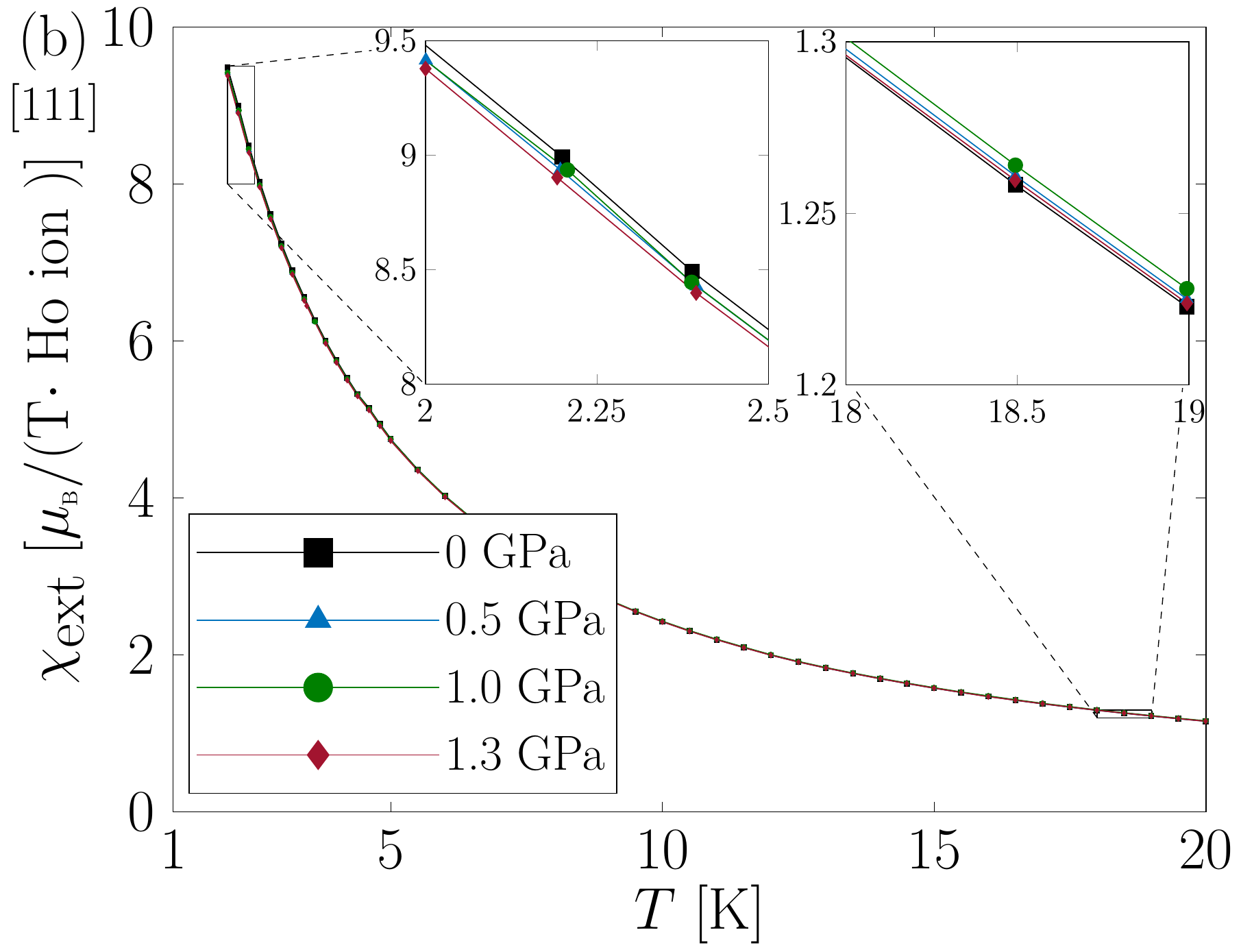}
	\end{subfigure}
	
    \caption{Susceptibility per holmium ion as function of temperature for field and pressure along the (a) $[001]$ and (b) $[111]$ crystalline directions. The field strength used to probe the susceptibility was $0.01\textup{ T}$. The insets show the change in susceptibility in the high and low temperature limit.} 
     \label{fig:rawData110andrawChi001111}
 \end{figure}

\subsection{Neutron scattering}\label{experimentalNeutronScatteringSection}
Magnetic neutron diffraction experiments were performed at the Institut Laue Langevin (ILL) using the polarized diffuse scattering instrument, D7\, \cite{D7}, with nominal incident wavelength $\lambda$ = $4.86\pm0.1$~\AA. We recorded neutron diffraction profiles for HTO in the $(h,-h,l)$ scattering plane under uniaxial pressure along the $[110]$ crystalline direction and with incident neutron polarization along the $[110]$ crystalline direction. For this experimental study, we cut our synthesized HTO crystals into cylinders of diameter $3.0\pm0.05\textup{ mm}$ and height $2.0\pm0.05\textup{ mm}$, still with major axis along $[110]$ but with larger diameter and smaller height than the crystals used in the magnetization measurement. Pressure was applied using a CuBe anvil-type pressure cell with a CuBe window of minimal thickness of $2.5\textup{ mm}$ for the entire scattering plane. Further details of the pressure cell are provided in appendix~\ref{Appendix:pressurecell} and a detailed overview of the pressure cell will be published elsewhere. The force was calibrated prior to the experiment using an in situ transducer for deviations experienced at cryogenic temperatures. The pressure was deduced from the force, allowing for some uncertainty via friction of the piston. We measured both the neutron spin-flip and non-spin-flip scattering as a function of the sample rotation about the major axis of the cylinder ($[110]$ crystalline direction). The data were corrected for detector and polarization analyzer efficiencies using standard samples of vanadium and amorphous silica, respectively \cite{D7}. Background measurements were performed at $300\textup{ K}$ with an Al sample, closely matched in dimensions to the HTO sample, to determine scattering from the pressure cell. The background was entirely symmetric in the sample holder rotation angle and the rotation-averaged background at ambient pressure was subtracted from the measured signal. 

\subsection{Experimental results}
Figures ~\ref{fig:rawData111.001} and \ref{fig:rawData110andrawChi001111} show the measured  magnetic moment and susceptibility at different pressures for the indicated directions. While the physics of spin ice in an applied field is a fascinating topic with features such as chain ordered states\cite{fennell05}, Kagome ice\cite{Higa03} and Kasteleyn transitions\cite{Fennell07}, our aim is to focus on the effects of pressure. We see a change in the magnetic moment on the order of a few percent under the application of pressure. For the $[001]$ measurement, Fig.~\ref{fig:rawData111.001}(a), the magnetic moment is monotonically reduced for all fields when pressure is applied. For the $[111]$ measurement, Fig.~\ref{fig:rawData111.001}(b), the magnetic moment is reduced both in the high and the low field limits, but for an intermediate field around $1-2\textup{ T}$ the magnetic moment increases as pressure is applied. For the $[110]$ measurement, shown in Fig.~\ref{fig:rawData111.001}(c), there is an overall decrease when pressure is applied, but the decrease is not monotonic at high fields, as it is for the other directions. There is also a large jump between the $p_{110}=0$ measurement and the $p_{110}>0$ measurements in the high and low field regions. We will discuss the implications of these features further in section~\ref{section:ResultMagnetizationAndSusceptibility}.

Figure~\ref{fig:rawData110andrawChi001111} shows the susceptibility, $\chi_\textup{ext}$, measured with field and pressure in the $[001]$ and $[111]$ crystalline directions. There is a noticeable change in the susceptibility when pressure is applied to the sample, but compared to the magnetic moments in Fig.~\ref{fig:rawData111.001} the effects are much less clear. For example, in our measurements using a small applied field of $0.01\textup{ T}$, the recorded change in the susceptibility is not generally monotonic with respect to the applied pressure.

\begin{figure}[h!]
    \centering
    \includegraphics[width=0.5\textwidth]{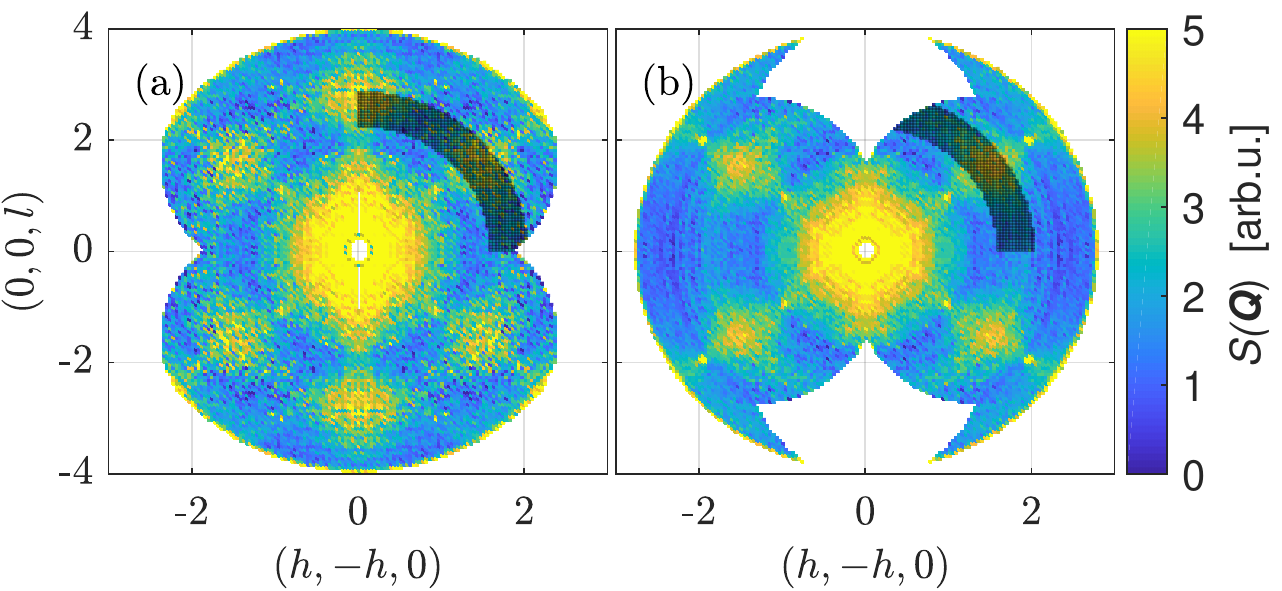}
    \caption{$[110]$ Spin-flip neutron scattering profile in the $(h,-h,l)$ plane at (a) $0\textup{ GPa}$ and (b)  $0.35\textup{ GPa}$\cite{dataThatWeTookOnILL_The110Direction}. The pressure is applied along the $[110]$ crystalline direction, perpendicular to the scattering plane. The data displayed have been symmetrized according to the crystal symmetry. Both measurements were conducted at $1.5\textup{ K}$. The shaded area marks the region which we integrate and analyze further in section~\ref{section:ResultNeutronScattering}. }
    \label{fig:NeutronScatteringExperiment110}
\end{figure}

Figure~\ref{fig:NeutronScatteringExperiment110} shows the spin-flip neutron scattering structure factor, $S(\textbf{Q})$, at ambient and at an applied pressure of $0.35\textup{ GPa}$ at $T=1.5\textup{ K}$\cite{dataThatWeTookOnILL_The110Direction}. The well-established  diffuse  scattering map\cite{HTO_J0.52proof_PhysRevLett.87.047205}  shows some of the defining features of the spin ice phenomenology including regions of intensity at (0,0,3),(3/2,3/2,3/2), and equivalent wave vectors, separated by pinch points at, for example, (0,0,2)\cite{Fennell, prr20}. In this study we focus on the effects of pressure on this scattering pattern. The noted pressure was applied at $300\textup{ K}$ and the subsequent pressure change as a result of thermal contraction could not be determined in situ. However, prior temperature dependent calibration measurements of the pressure cell provides an indication that the noted pressure is approximately correct. A uniaxial pressure cell with in situ pressure determination for diffuse neutron scattering and polarization analysis is under development. In Fig.
~\ref{fig:NeutronScatteringExperiment110}(a,b), the measurements are taken with two different samples, with equivalent sample dimensions, as the first sample broke when pressure was applied. The samples were cut from the same single crystal. We note a change in the intensity of the diagonal satellite peaks ($h\approx \pm1.5, l\approx\pm 1.5$) when pressure is applied. To better demonstrate these subtle changes, we have integrated the shaded regions over $Q$, see Fig
~\ref{fig:NeutronScatteringExperiment110}, and fitted a Gaussian curve to the integrated data. We find that the intensity of the diagonal satellite peaks increases by about $4\%$ when pressure is applied, and discuss this further in section~\ref{section:ResultNeutronScattering}.

\section{Theoretical modeling}
\subsection{Model and simulation method}
A theoretical model for the evolution of the interactions in classical spin ice under uniaxial pressure was proposed in an earlier investigation\cite{OurArticle}. We adapt this model to the current measurements of HTO and refer to it as the dipolar spin-ice pressure model (DSPM). The DSPM is an extension of the standard dipolar spin-ice model (s-DSM)\cite{MelkoGingras}. The Hamiltonian for the DSPM for classical unit Ising spins $\textbf{S}_i$ on the pyrochlore lattice is defined as 
\begin{equation}\label{Hamiltonian}
\begin{split}
\mathcal{H}=&\sum_{\langle i,j\rangle}J(i,j) \, \mathbf{S}_i\cdot\mathbf{S}_j\\+
&Da^3\sum_{i<j}\left(\frac{\mathbf{S}_i\cdot\mathbf{S}_j}{|\mathbf{r}_{ij}|^3}-3\frac{\left(\mathbf{S}_i\cdot\mathbf{r}_{ij}\right)\left(\mathbf{S}_j\cdot\mathbf{r}_{ij}\right)}{|\mathbf{r}_{ij}|^5}  \right).
\end{split}
\end{equation}
\begin{figure}[h!]
	\centering
	\includegraphics[width=1\linewidth]{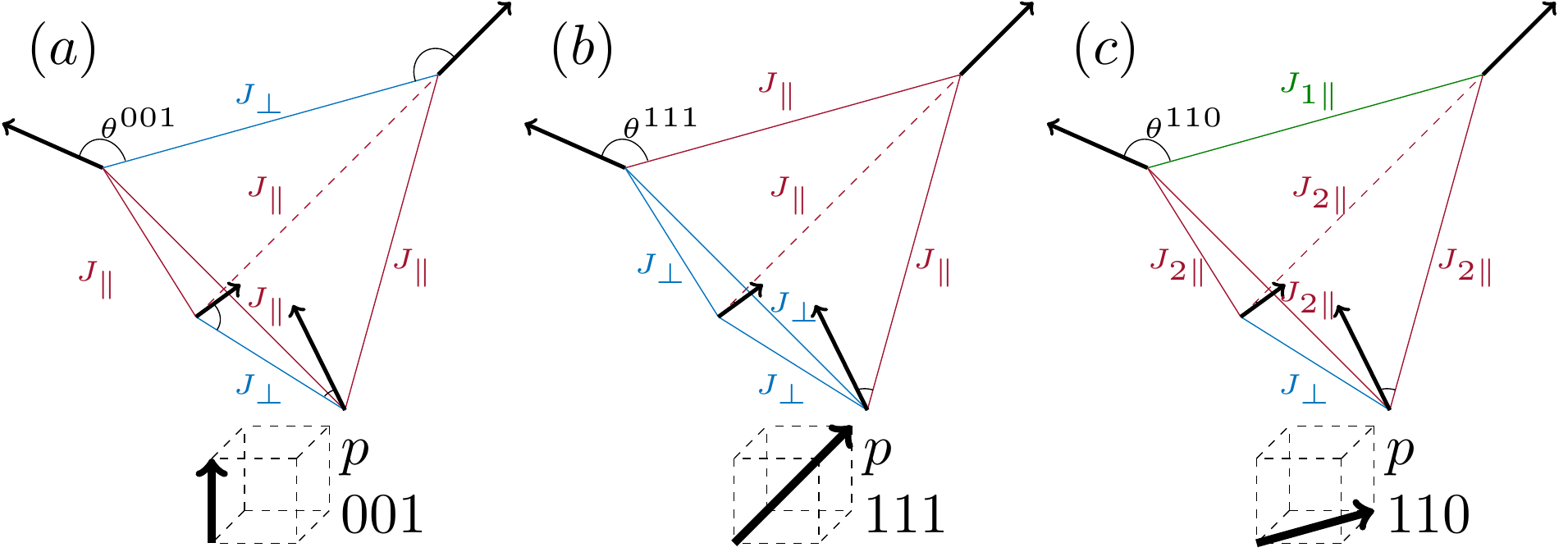}
	\caption{Illustrations of the DSPM\cite{OurArticle}.  Pressure is applied along the (a)$[001]$, (b)$[111]$, (c)$[110]$ direction and the colors indicate the different exchange interactions $J_\parallel (\textup{red}), J_{1\parallel} (\textup{green}), J_{2\parallel} (\textup{red}), J_\perp$(blue). The angles of the Ising moments $\theta^{001},\theta^{111},\theta^{110}$ are assumed to change under application of pressure so that spins keep pointing toward the center of the compressed tetrahedron. }
	\label{fig:model}
\end{figure}

The strength of the dipolar interactions is taken to be  $D=1.41\textup{ K}$\cite{Original_D1.41_PhysRevLett.84.3430}, $a$ is the nearest neighbor distance, $\textbf{r}_i$ is the position of spin $\textbf{S}_i$ and $\textbf{r}_{ij}=\textbf{r}_i-\textbf{r}_j$. The angled brackets in the first sum denotes summation over nearest neighbors. The antiferromagnetic nearest neighbor interaction $J(i,j)$ is caused by oxygen-mediated superexchange. The value of $J$ at ambient pressure is set to $J_{p=0}=1.56\textup{ K}$\cite{HTO_J0.52proof_PhysRevLett.87.047205}. In the DSPM it is assumed that $J_{\perp}=J_{p=0}=1.56 \textup{ K}$ if the neighbors $i$ and $j$ are in a lattice plane perpendicular to the direction of applied pressure and $J(i,j)=J_{\parallel}$ if they are not, ($J_{1\parallel}$,$J_{2\parallel}$ for pressure in the $[110]$ direction as there is less symmetry) see Fig.~\ref{fig:model}. The motivation for this assumption is that the distance between neighboring ions, and hence the exchange integral, will change when pressure is applied. Compression of the crystal is modeled with zero Poisson ratio, which is also the case in the current experimental setup for the measurements of the magnetic moment\cite{PaperPascaleWantedToCiteForZeroPoissonRatio_Vol6921642000}. As the lattice is compressed, the local Ising axis of the spins is modeled to keep pointing towards the centers of the compressed tetrahedra. The dipolar interaction becomes stronger along the direction of compression, growing as $|\textbf{r}_{ij}|^{-3}$ with decreasing distance. We define the lattice compression $\kappa$ as the relative length contraction along the direction of pressure. At high fields, the saturated magnetic moment is dependent only on $\kappa$, due to the change in angles of the Ising moments. The high field saturated value of the magnetic moment therefore sets the lattice compression, and we use the low field data to determine exchange interactions.

At this point, a discussion of the crystal field effects in HTO is necessary. The ionic state of the Ho$^{3+}$ ion depends on the local electric fields of the surrounding ion. Shifting their position by applying pressure could affect the ionic ground state, on which our present analysis is based. Under ambient pressure, the electronic $^5I_8$ ground state of a single Ho$^{3+}$ is lifted by the trigonal field of the surrounding oxygen atoms. Fitting the crystal field parameters to susceptibility and inelastic neutron measurements yields an almost pure $|J, M_J\rangle=|8,\pm8\rangle$ non-Kramers ground state doublet, with an excitation gap of more than 200K, and an almost full magnetic moment of about 10 $\mu_B$\cite{jana00,rosen00}. These are the two crucial points for our study. Although the ground state doublet in HTO is not protected by Kramers' theorem, like it is for DTO, it is unlikely that the small shift in the ionic positions which we induce by applying pressure would significantly affect the nature of the ground state doublet leading to our current Ising description. Considering a perturbation theory approach, it is not unlikely that the magnetic moment of the Holmium ion could shift slightly due to crystal field effects, since this is a first-order effect. However, the changes in the ground state itself are expected to be very small, since this second-order effect is suppressed by the very significant energy gap, and by the fact that the ground state has a very small overlap with states other than $|J, M_J\rangle=|8,\pm8\rangle$.To completely rule out crystal field effects would require  a significant experimental and theoretical undertaking, but we believe the above perturbative argument to be strong enough to proceed with our analysis for HTO. We also note that pressure induced changes in the crystal field parameters in general can be expected to be more significant. For example, in the related compound  Tb\textsubscript{2}Ti\textsubscript{2}O\textsubscript{7} (TTO), the energy gap is ten times smaller than in HTO, and the admixture of the ground state doublet is significant\cite{boothroyd15}.  This also makes  TTO an interesting case study for application of pressure, but the theoretical analysis would require a very different approach from the one presented here.

Monte Carlo (MC) simulations using the Metropolis–Hastings algorithm and single spin flip updates are used to investigate a number of different system sizes with periodic boundary conditions. Since the lowest temperature used in the experiment was $1.5\textup{ K}$, monopole excitations are still prevalent and loop flips\cite{Melko01} are not needed. We use the 16-particle standard cubic unit cell. All supercells are cubic of size $L^3$ unit cells, $L  \in [1,...,8]$. Ewald summation is used to effectively account for the long-range conditionally convergent dipolar contributions\cite{frenkel2001understanding}.

\subsection{Theoretical misalignment effects}\label{subsection_Magnetization}

The DSPM  describes the changes in the magnetic moment observed in measurements\cite{MITO} of DTO under uniaxial pressure\cite{OurArticle} quite well. Particularly when pressure and field is applied along the $[001]$ or $[111]$ crystalline directions the model works very well. However, when pressure and field is applied in the $[110]$ crystalline direction for DTO, anomalous behavior is observed which cannot be explained by the DSPM\cite{MITO,OurArticle}, since  the pressure-induced change in the magnetic moment does not saturate at high fields\cite{MITO,OurArticle}. 

In this study, we attempt to explain this previously noted discrepancy for the measurement in the $[110]$ crystalline direction and speculate that it can be an experimental artefact due to misalignment of the crystal. Next, we outline a brief discussion of what effects might be expected from crystal misalignment. 

When measuring the magnetic moment of a single crystal under uniaxial pressure, there are three quantities that need to be aligned: the crystalline direction, the direction of the uniaxial pressure and the direction of the magnetic field. 

In order to investigate the influence of misalignment, we consider the possibility that the crystalline direction matches the direction of pressure perfectly, and that the magnetic field is misaligned with respect to the direction of pressure. This  assumption minimizes the number of new parameters introduced to the theoretical model, and we believe that it still captures the essential effects of misalignment. In the experiment, on the other hand, we may expect the misalignment between the crystal direction and the applied pressure to be more significant. Since the crystal is surrounded by more compressible Stycast it is also possible that this misalignment may change during a series of compressions. See appendix~\ref{Appendix:misalignment} for further discussion.

\begin{figure}[h!]
    \centering
    \includegraphics[width=0.8\linewidth]{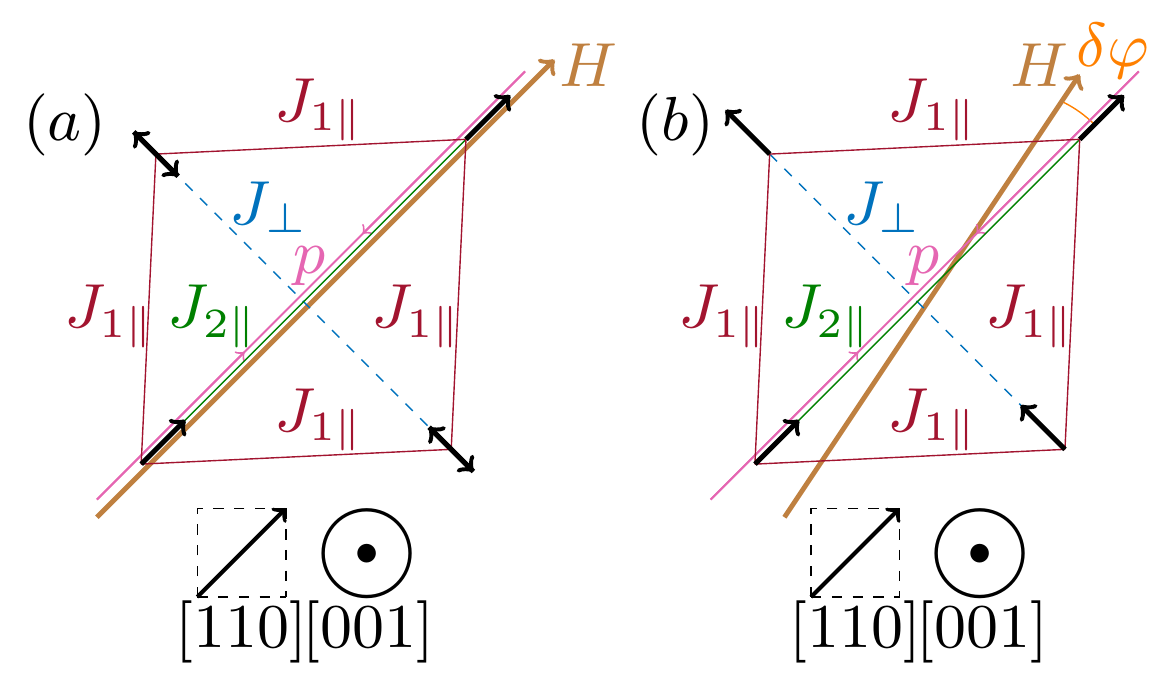}
    \caption{View along $[001]$ of a single tetrahedron with pressure along $[110]$. (a) Perfect alignment. (b) Field misalignment perpendicular to $[001]$. We assume that the misalignment angle $\delta\varphi$ can vary with application of pressure. For a field in this direction there are two degenerate ground states for perfect alignment (double headed arrows in (a)), and only one ground state when the field is misaligned (b). We denote the misalignment at ambient and applied pressure by $\delta\varphi^0$ and $\delta\varphi^p$ respectively.}
    \label{fig:misalignment}
\end{figure}

Misalignment has particularly strong effects when the measurement is conducted with pressure along the $[110]$ direction\cite{MisalignmentReference2001Fukazawa}. This is because half of the spins in the material have a local Ising axis perpendicular to the $[110]$ direction, so that at high fields, already a small misalignment will strongly affect the dynamics of these. We therefore adjust the applied field so that it is misaligned by an angle $\delta\varphi$ with respect to the $[110]$ direction and remains perpendicular to $[001]$, as this is the simplest way for the field to couple to the previously unaffected spins. Fig.~\ref{fig:misalignment} illustrates the scenario. The magnetic moment is measured along the direction of the field, and when the field is misaligned we start to observe the dynamics of the previously perpendicular spins. This is not the case when we measure along the $[001]$ and $[111]$ crystalline directions, since in that case, all spins have already a large component along the field. Hence we expect that the high field behavior will be less affected by misalignment for measurements in the $[001]$ and $[111]$ crystalline directions. Our simulations confirm that this is indeed the case. The saturation fields for the pressure induced changes in magnetic moment in the $[001]$ and $[111]$ measurements are not affected by misalignment, in contrast to the $[110]$ direction. The effects of misalignment in the $[110]$ direction are presented in section~\ref{section:ResultMagnetizationAndSusceptibility} and the key result is then shown in Fig.~\ref{fig:delta001.111.110}(c).

\subsection{Demagnetizing corrections}
In order to account for the macroscopic boundary effects, we perform a demagnetizing transformation. The samples used were cylindrical with an aspect ratio of $\gamma \equiv\frac{\text{height}}{\text{diameter}}=1.50\pm 0.05$ at ambient pressure. However, as the the sample is compressed, the aspect ratio decreases. By fitting $\kappa$ to the change in saturated magnetic moment we determine $\gamma$. An important aspect, often overlooked, is that the demagnetization factor $N$ is a function of both $\gamma$ and the internal volume susceptibility $\chi^V_\textup{int}$, $N=N(\gamma,\chi_\textup{int}
^V)$ as has been recently calculated and verified\cite{CHEN2006135,tweng17}. In high-susceptibility materials ($\chi_\textup{int}>1$ ), such as HTO, the susceptibility dependence of $N$ is very significant and we include it in the analysis. We perform a demagnetization transformation according to
\begin{equation}\label{eq:demagtrans}
\begin{split}
H_\textup{ext}&=H_\textup{int}+N(\gamma,\chi_\textup{int}^V) M^V,\\
\chi_\textup{ext}^V&=\frac{ \chi_\textup{int}^V}{1+N(\gamma,\chi_\textup{int}^V)\chi_\textup{int}^V},
\end{split}
\end{equation} 
where the intensive quantities $M^V$ and $\chi^V$ are the magnetization and volume susceptibility respectively. It is important to differentiate these from the external extensive quantities $M=M^VV(\gamma(p))$ and $\chi_\textup{ext}=\chi_\textup{ext}^VV(\gamma(p))$, measured in the experiment. The Monte Carlo calculations, performed using periodic Ewald boundary conditions, yield internal quantities.

\begin{figure}[h!]
    \centering
    \includegraphics[width=0.85\linewidth]{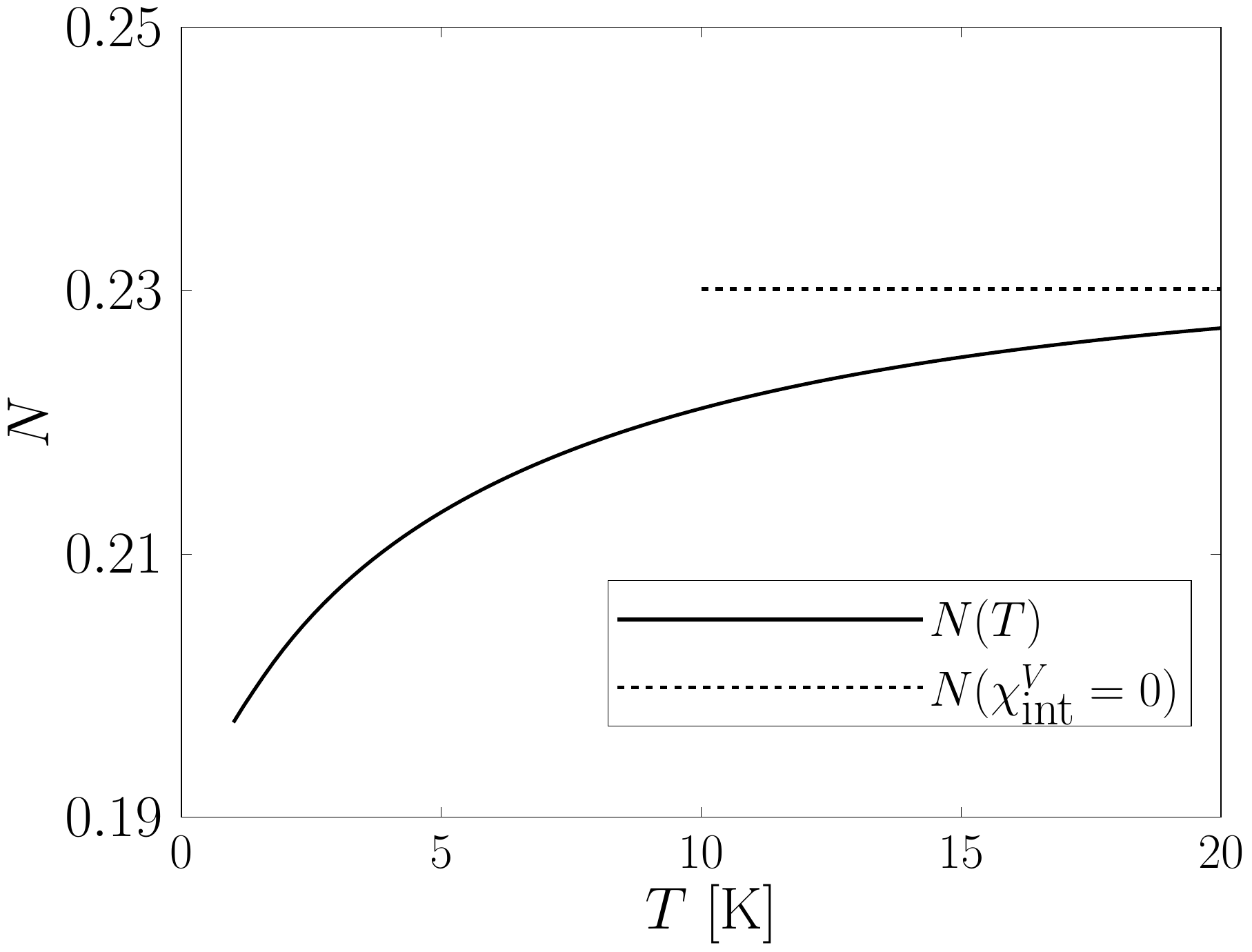}
    \caption{Demagnetizing factor $N$ as a function of $T$ at zero field for a HTO cylinder of aspect ratio $\gamma=1.5$.}
    \label{fig:N}
\end{figure}

Since the susceptibility of classical spin ice is isotropic the susceptibility dependence of $N$ calculated in Ref.~\onlinecite{CHEN2006135} can be used for HTO, as demonstrated in Ref.~\onlinecite{tweng17}. Using the tabulated values of $N(\gamma,\chi_\textup{int}^V)$ we perform an interpolation using cubic splines\cite{CHEN2006135}. As an example, we show the temperature dependence of $N$ for an HTO  cylinder of aspect ratio $\gamma=1.5$ in Fig.~\ref{fig:N}, where we use the experimentally measured intrinsic susceptibility for HTO\cite{bovo13} at zero field to determine the temperature dependence.


\subsection{Calculation of susceptibility}
For the susceptibility measurements, we define the relative change in the susceptibility with respect to the applied pressure as $\Delta\chi_\textup{ext}/\chi_\textup{ext} \equiv	(\chi_{\textup{ext}}^p-\chi^0_{\textup{ext}})\chi^0_{\textup{ext}}$.
We measured the relative change $\Delta\chi_\textup{ext}/\chi_\textup{ext}$ as a function of temperature. This quantity is highly sensitive to changes in $J$ and $\kappa$, and in order to gain some physical insight, we  consider the high-temperature limit, $T\to\infty$. In this non-interacting limit the relative change depends only on $\kappa$. For the crystalline directions along which measurements were performed we find that
\begin{equation}\label{eq:dChi/chi_analytic_highT}
\begin{split}
    \lim_{T\to\infty}  \frac{\Delta \chi^{111}_\textup{ext}}{\chi^{111}_\textup{ext}}&=\frac{9\,{\left(1-\kappa\right)}^2}{4\,{\left(1-\kappa\right)}^2+32}-\frac{1}{4},\\
    \lim_{T\to\infty}  \frac{\Delta \chi^{001}_\textup{ext}}{\chi^{001}_\textup{ext}}&=\frac{3(1-\kappa)^2}{2+(1-\kappa)^2}-1,
\end{split}
\end{equation}
where the superscripts denote the crystalline direction of pressure and field.

From these formulae, we see that in both directions, the susceptibility at high temperature is reduced under application of pressure. This is a purely geometrical effect stemming from the fact that the Ising moments tilt away from the axis of pressure when the lattice is compressed.

\subsection{Calculation of magnetic structure factor}
We calculate theoretical predictions for the spin-flip magnetic structure factor according to
\begin{equation}\label{Eq.SFstructureFactor}
\begin{split}
&S(\textbf{Q})=\frac{[f(|\textbf{Q}|)]^2}{\mathcal{N}}\\
&\times\sum_{ij}\left\langle \textbf{S}^{\perp}_i\cdot \textbf{S}^{\perp}_j-(\textbf{S}_i\cdot \textbf{P})(\textbf{S}_j \cdot \textbf{P})\right\rangle e^{i\textbf{Q}\cdot \textbf{r}_{ij}},
\end{split}
\end{equation}
where the scattering wavevector is denoted by $\textbf{Q}$, the normalized polarization vector of the incident neutron beam is given by $\textbf{P}\perp\textbf{Q}$. The component of the spin perpendicular to the wave vector is defined as  ${\textbf{S}_i^\perp=\textbf{S}_i-\textbf{S}_i\cdot\textbf{Q}/|\textbf{Q}|^2\textbf{Q}}$. $\mathcal{N}$ is the number of particles in the supercell and $f(|\textbf{Q}|)$ is the magnetic form factor for Ho\textsuperscript{3+}. The angled brackets $\langle\cdots\rangle$ denote the thermal average which is calculated using the MC method.

\begin{figure}[h!]
	\begin{subfigure}[t]{0.5\textwidth}
		\centering
		\includegraphics[width=0.85\linewidth]{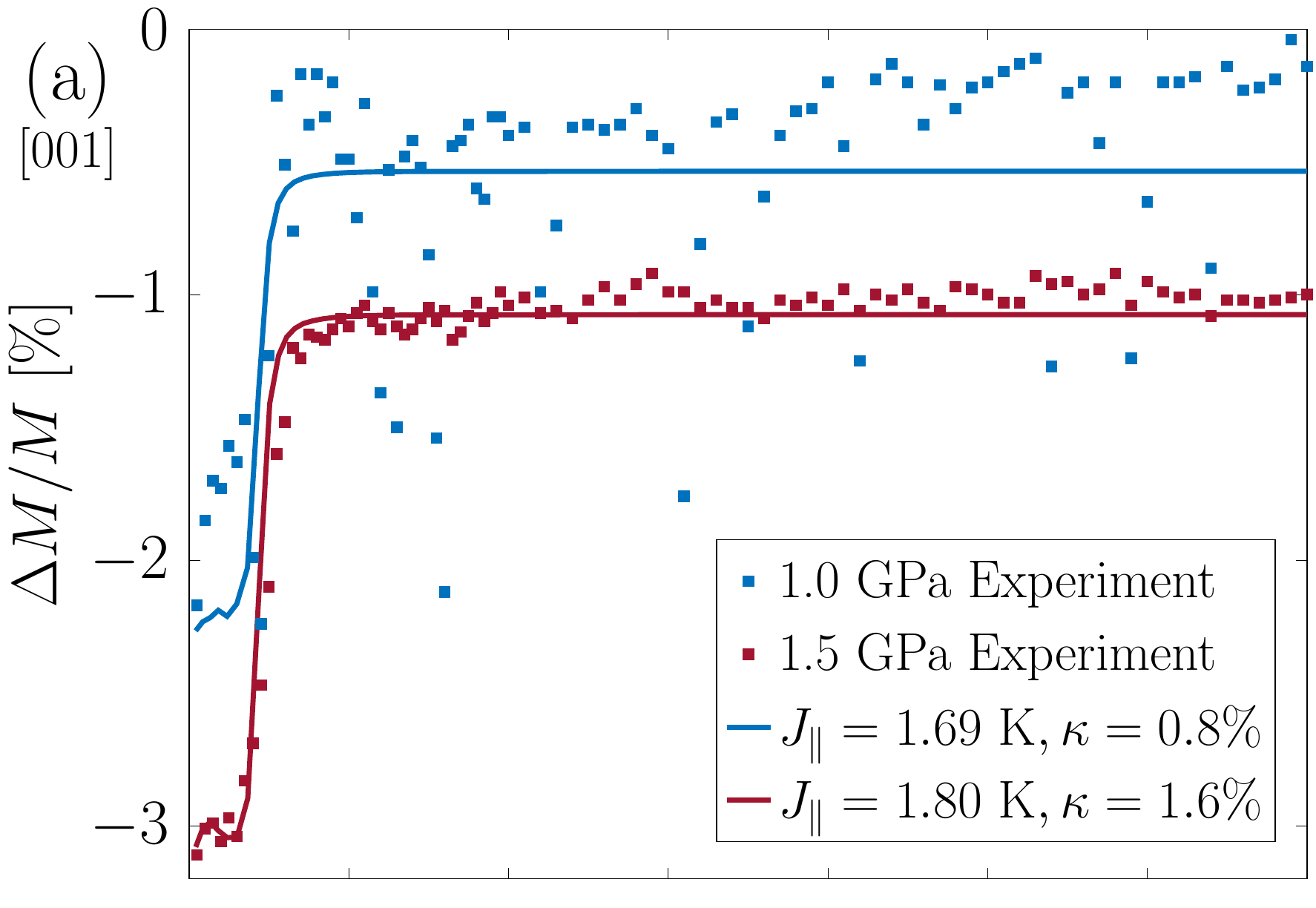}
	\end{subfigure}
	\begin{subfigure}[t]{0.5\textwidth}
		\centering
		\includegraphics[width=0.85\linewidth]{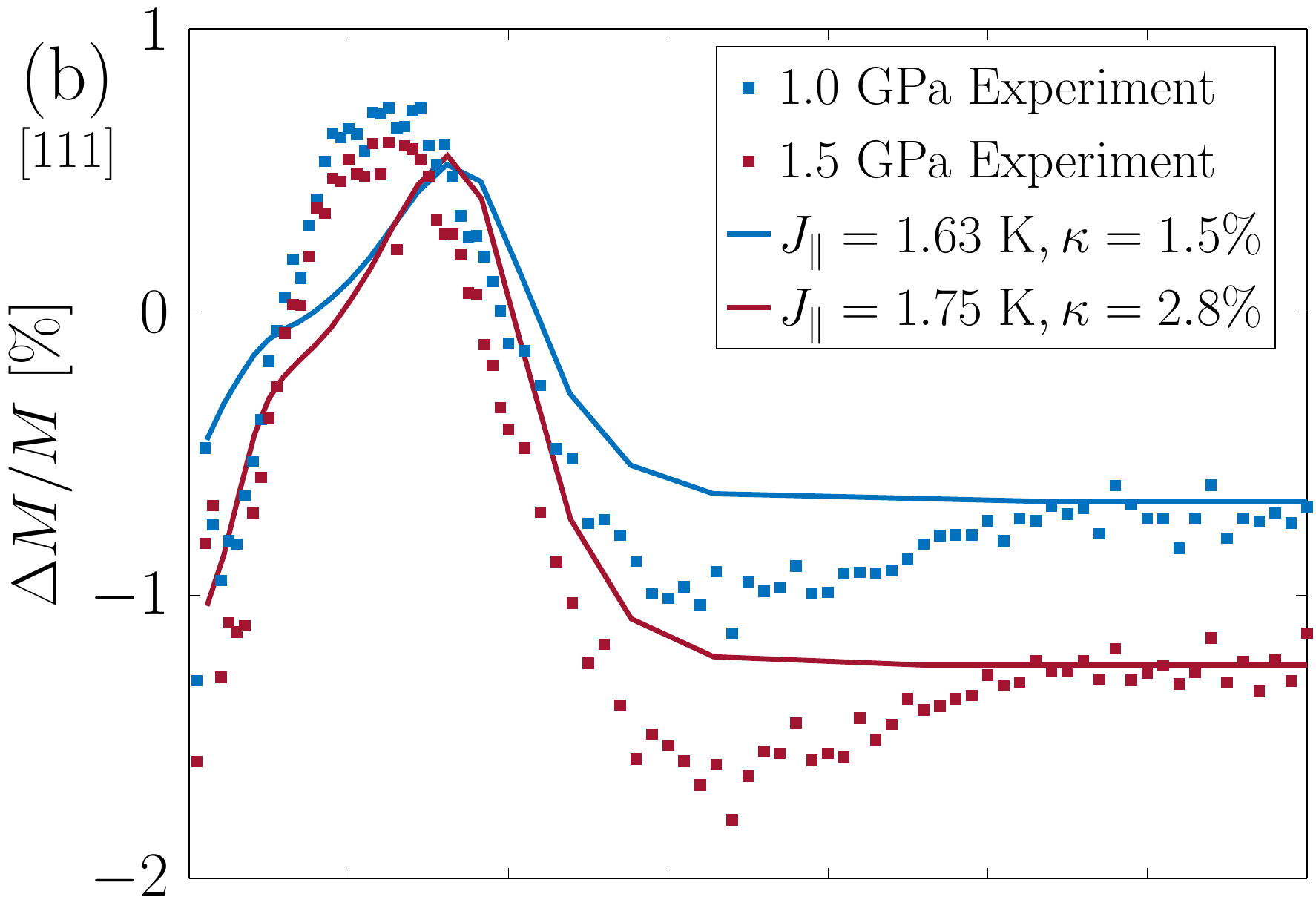}
	\end{subfigure}
	\begin{subfigure}{0.5\textwidth}
	\centering
\hspace{-0.15mm}
	\includegraphics[width=0.855\linewidth]{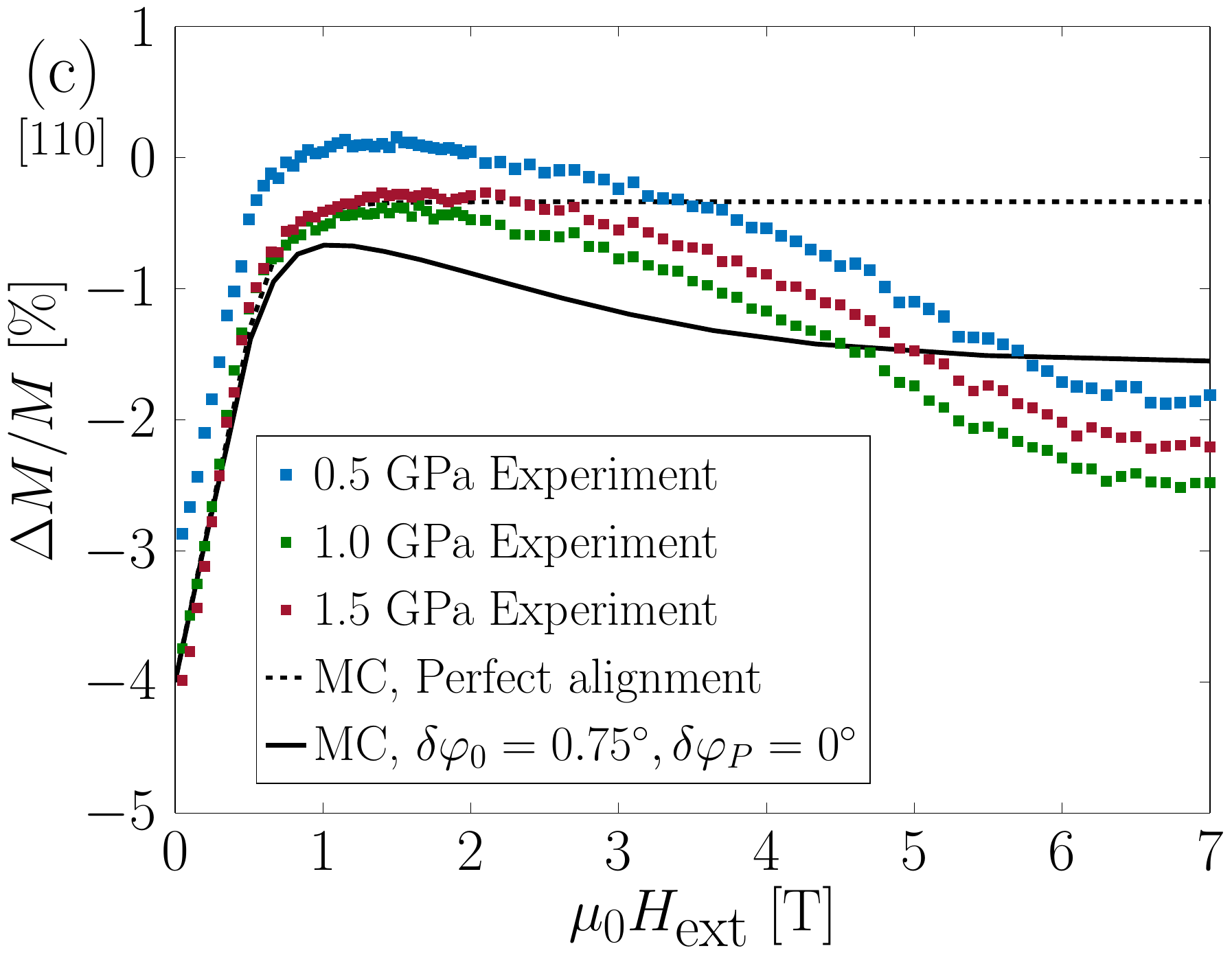}
    \end{subfigure}
	\caption{Relative change in sample magnetic moment, $\Delta M/M=(M^p-M^0)/M^0$, as a function of magnetic field at sample temperature $1.83\textup{ K}$.  The pressure and field are along the (a)$[001]$, (b)$[111]$, (c)$[110]$ crystalline direction. MC results are drawn as solid lines and performed on cubic systems of $8192$ particles, $L=8$.  For the $[110]$ measurement (c) we show the model prediction for $J_{1\parallel}=J_{2\parallel}=1.2\textup{ K}, \kappa=1\%$ both for perfect alignment(dashed line) and for a small misalignment when pressure was applied(solid line).}
	\label{fig:delta001.111.110}
\end{figure}

\section{Analysis and discussion}\label{Section:resultSection}

\subsection{Magnetization and susceptibility}\label{section:ResultMagnetizationAndSusceptibility}

In Fig.~\ref{fig:delta001.111.110} we show the relative change in sample magnetic moment, $\Delta M/M=(M^p-M^0)/M^0$, as function of applied magnetic field. The solid lines show the best theoretical fit for the different directions of pressure and field. For the $[001]$ and $[111]$ measurements, shown in Fig.~\ref{fig:delta001.111.110}(a,b), $\kappa$ has been set so that the change in calculated saturation magnetic moment agrees with that of the experiment. With this constraint, there is only one free parameter, $J_\parallel$, which we determine from the best fit for each case. We find that for the $[001]$ and $[111]$ measurement, $J_\parallel$ increases monotonically with pressure. The change in $J_\parallel$  is similar in both directions and the model accommodates the basic features in the experimental data: the upturn at $1.5 \textup{ T}$ in the $[111]$ measurement as well as the drop at $0.5\textup{ T}$ in the $[001]$ direction, upon decreasing field. For the $[001]$ measurement we have excellent agreement between theory and experiment. We note that the observed decrease in the magnetic moment for HTO at low fields contrasts to measurements made on DTO, where an increase of $\Delta M/M=4\%$ was observed\cite{MITO}. Although $J_\parallel$ and $\kappa$ increase similarly for both materials, the ratio $J/D$ is smaller for HTO, and the effects of the dipolar interactions dominate, leading to a decrease in the magnetic moment. In the $[111]$ direction  the experimentally observed minimum centered at $\mu_0H_\textup{ext}= 3.5 \textup{ T}$ is not  accounted for in the model, and the shape of the maximum around $1.8\textup{ T}$ differs, but the model captures the qualitative experimental features. Furthermore, in all measurements, the change in magnetic moment is strongly dependent on the field. This is due to competition between different effects. At large fields, the moment is saturated and since spins tilt away from the direction of pressure, it decreases. At low fields, the thermal fluctuations determine the magnetic moment. In the case of HTO, the magnetization decreases when pressure is applied, while for DTO in the $[001]$ direction, the preference for a ferromagnetic ground state is indicated by an upturn in the magnetization.

Figure~\ref{fig:delta001.111.110}(c) shows the change in magnetic moment for the $[110]$ measurement. We note that the change has hardly saturated as a function of field. Seemingly, we need a field of about $7\textup{ T}$ to saturate the change in the $[110]$ direction, higher than the $5.5\textup{ T}$ needed for the $[111]$ direction. This feature is not possible to reproduce in the DSPM in which the field needed for $[111]$ saturation is higher than that needed for $[110]$ saturation, since the saturated $[111]$ state requires that we break the two-in two-out ground state of spin ice\cite{harris97}.

As discussed in section~\ref{subsection_Magnetization}, we speculate that the high-field unsaturated behavior can be due to misalignment of the crystal, since measurements are particularly sensitive in the $[110]$ direction. In Fig.~\ref{fig:delta001.111.110}(c) we illustrate the influence of misalignment. We first fit a curve with $\kappa=1\%$, as an estimate of the compression based on the other directions at $1\textup{ GPa}$ ($0.8\%$ and $1.5\%$ for $[001]$ and $[111]$ respectively). In the DSPM, the change in magnetic moment will saturate at about $1\textup{ T}$ (same order of magnitude as in the $[001]$ case) regardless of the parameters used. For simplicity we assume that $J_{1\parallel}=J_{2\parallel}=J_\parallel$ which gives a rough value of $J_\parallel=1.2\textup{ K}$ when we fit this single parameter to match the sub $1\textup{ T}$ experimental results. This DSPM fit is shown as the dashed line in Fig.~\ref{fig:delta001.111.110}(c). For all fields above $1\textup{ T}$, the DSPM gives a constant saturated change in magnetic moment dependent only on $\kappa$. Clearly this is not what we see in the experiment, where the change keeps varying up to fields of about $7\textup{ T}$.

If we introduce crystal misalignment according to section~\ref{subsection_Magnetization}, we can partially reproduce the high-field trend observed in the measurement. Within this framework, we have two additional parameters $\delta\varphi^0, \delta\varphi^p$ as discussed in section~\ref{subsection_Magnetization} for the misalignment of $H$ with respect to the $[110]$ direction, at zero($0$) and applied pressure($p$) respectively. To get a decrease in magnetic moment when pressure is applied, we find it necessary that the field couples stronger to the perpendicular spins in the reference measurement. Hence, to keep the model as simple as possible, we set $\delta\varphi^p=0^\circ$. We then adjust $\delta\varphi^0$ in order to fit the experimental curve above $1\textup{ T}$. We find that $\delta\varphi^0=0.75^\circ$ gives the best fit. We do not get the exact same features as those observed, but we do demonstrate that the decrease in magnetization above $1 \textup{ T}$ could be due to misalignment. For further discussion of more elaborate misalignment models under pressure we refer to appendix~\ref{Appendix:misalignment}.

\begin{figure}[h!]
    \begin{subfigure}[t]{0.5\textwidth}
		\centering
		 \includegraphics[width=0.83\linewidth]{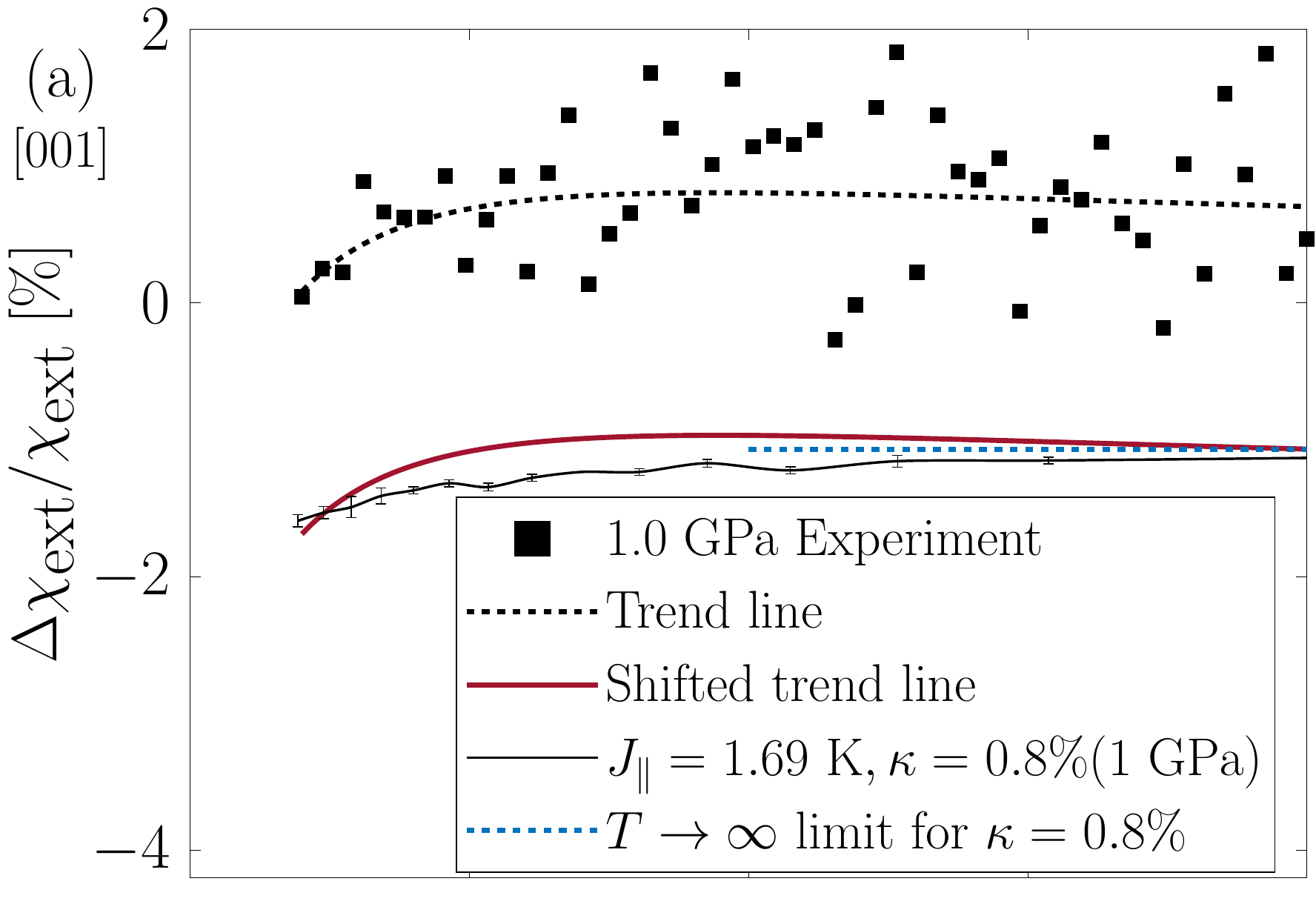}
	\end{subfigure}
	\begin{subfigure}[t]{0.5\textwidth}
    	\centering
    	\hspace{0.2mm}
	    \includegraphics[width=0.85\linewidth]{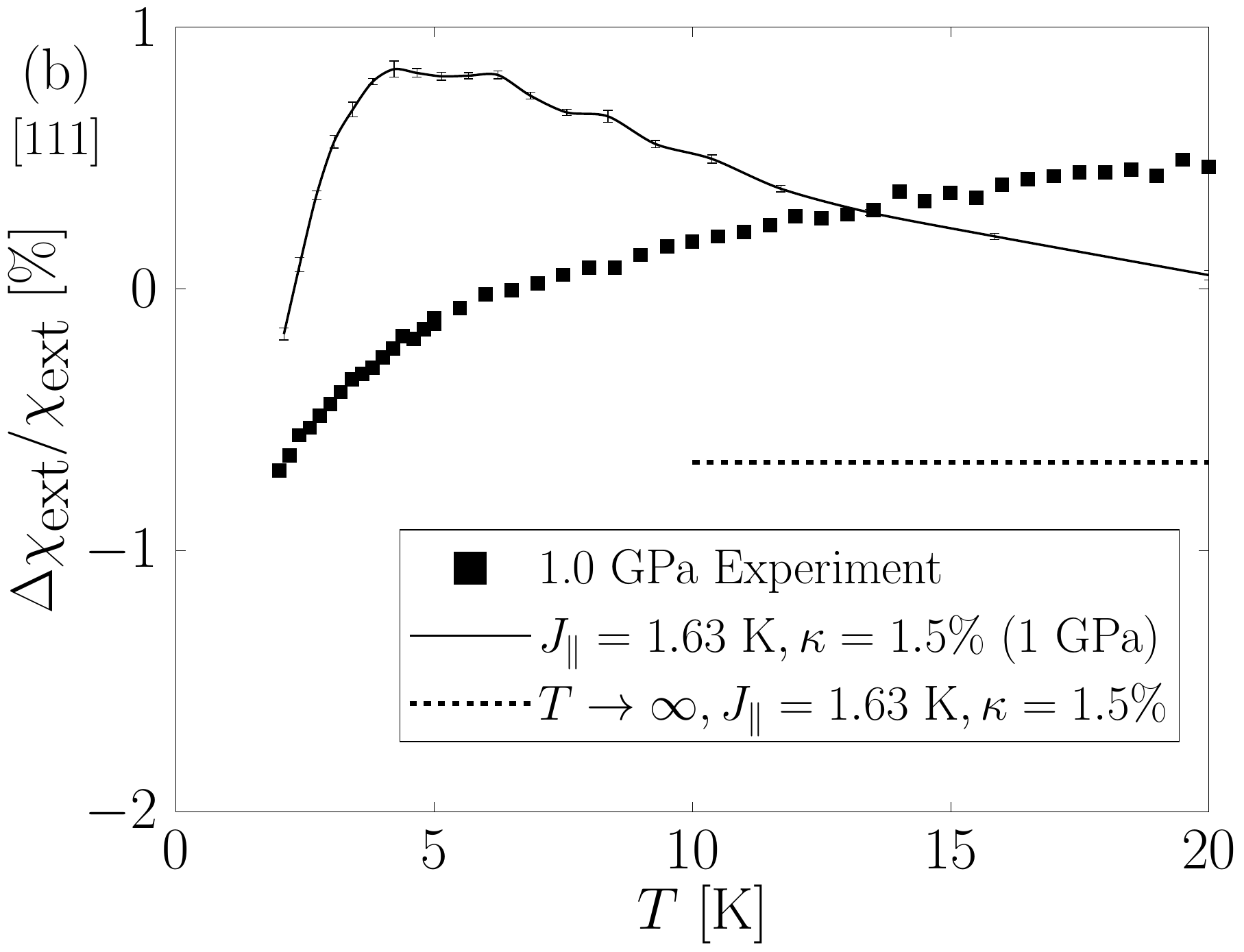}
	\end{subfigure}
	\caption{Relative change in external susceptibility for pressure and field along the (a) $[001]$ (b) $[111]$, both for probing external field of strength $0.01\textup{ T}$. Solid lines show the theoretical MC prediction from the parameters fitted from the magnetic moment measurements at $1.0\textup{ GPa}$. Dashed lines indicate the $T\rightarrow\infty$ analytical limiting value, Eq.~(\ref{eq:dChi/chi_analytic_highT}). MC simulation for $2000$ particles, $L=5$.}
	\label{fig:deltaChi}
\end{figure}

Finally, we use our obtained values of $J_\parallel$ and $\kappa$ to predict the change in susceptibility as a function of temperature. In Fig.~\ref{fig:deltaChi} we show the measured relative change in external susceptibility plotted against temperature for an applied pressure of $1\textup{ GPa}$. The theoretical predictions are shown for the values derived from the $1.0\textup{ GPa}$ $[001]$ and $[111]$ magnetic moment fits, Fig.~\ref{fig:delta001.111.110}(a,b), respectively. In contrast to the measurements of the magnetic moment, the susceptibility measurements are more challenging due to the low field used and a higher sensitivity to uncertainties in the demagnetizing factors.

Due to the fluctuations in the  $[001]$ measurement, Fig.~\ref{fig:deltaChi}(a),  we include a trend line by fitting rational polynomials to the susceptibility. The qualitative curve shape is the same for both theory and experiment, but they differ  by an overall shift along the vertical axis. At the lowest temperature $T=1.83\textup{ K}$ we expect that the relative change in susceptibility should coincide with the $H_\textup{ext}\to0$ limit in the relative change in the magnetic moment Fig.~\ref{fig:delta001.111.110}(a), which for $1\textup{ GPa}$ is $-2\%$. There is therefore a discrepancy between the two experimental measurements. From theory we also expect that the susceptibility should be reduced at high temperature when pressure is applied, see Eq.~(\ref{eq:dChi/chi_analytic_highT}). Since the susceptibility measurements in Fig.
~\ref{fig:rawData110andrawChi001111} were nonmonotonic in pressure we suspect a systematic error in the experimental data and note that if we shift the susceptibility measurement performed under pressure by $-2\%$ the data matches the theoretical prediction as well as the magnetization measurement performed at $T=1.83\textup{ K}$.

For the $[111]$ direction there is a more significant mismatch between theory and experiment. In particular, theory would predict a maximum in $\Delta\chi/\chi$ near $5\textup{ K}$, which is not present in the experiment. We would also expect from theory that the relative change should saturate to a lower value. Indeed, with the assumption of Ising spins it directly follows from Eq.~(\ref{eq:dChi/chi_analytic_highT}) that $\Delta\chi/\chi$ must reach a negative value at high temperature. The discrepancy is most likely due to  insufficient accuracy in the present susceptibility measurements.

\subsection{Neutron scattering}\label{section:ResultNeutronScattering}
\begin{figure}[h!]
    \centering
    \includegraphics[width=0.85\linewidth]{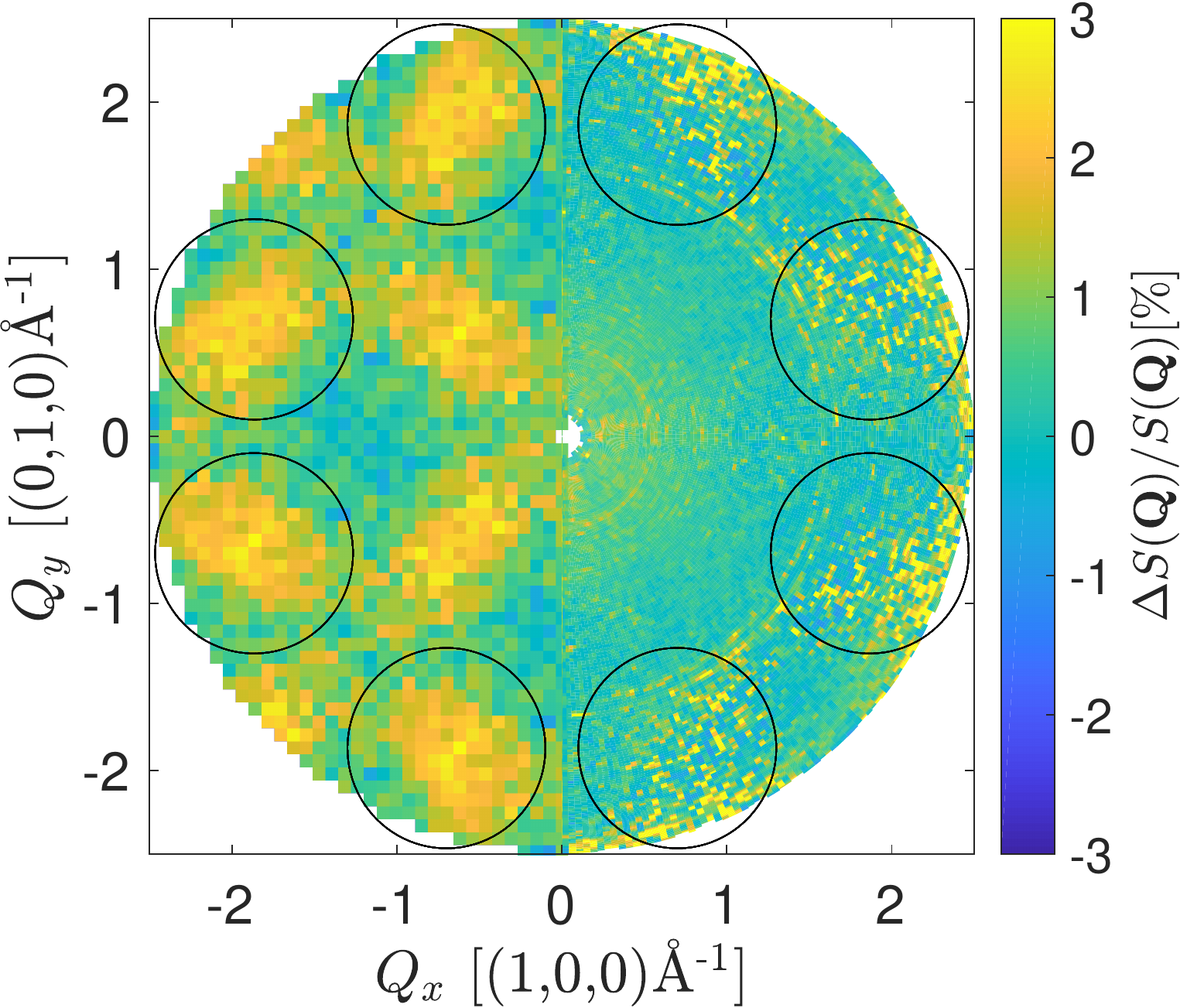}
    \caption{(left) Relative change in the $[001]$ spin-flip neutron scattering profile predicted from parameters found in the measurement of the magnetic moment, Fig.~\ref{fig:delta001.111.110}, at $1.5\text{ GPa}$ along $[001]$ and $1.5\textup{ K}$  ($J_\parallel=1.80,\kappa=1.6\%$). (right) Relative change in experimental $[001]$ spin-flip neutron scattering data\cite{OurArticle,dataThatWeTookOnILL_The001Direction} for HTO at $2.2\textup{ GPa}$ pressure along $[001]$, $1.5\textup{ K}$. Circles mark regions of increased intensity and are plotted at symmetry equivalent regions in the theoretical prediction. In previous investigations\cite{OurArticle}, the DSPM parameters were estimated based directly on this experimental scattering profile. The current measurements of the magnetic moment show consistency with this estimate.}
    \label{fig:dSQ001}
\end{figure}

The parameters extrapolated from the measurements of the magnetic moment can be used to compute the thermal spin-spin correlation function and hence the magnetic structure factor, Eq.~(\ref{Eq.SFstructureFactor}) within the DSPM. For pressure along the $[001]$ crystalline axis theoretical  and experimental results match well, see Fig.~\ref{fig:delta001.111.110}(a), and we use the parameters from this fit to predict the $[001]$ spin-flip structure factor under pressure. The resulting change in $S(\textbf{Q})$, $\Delta S(\textbf{Q})/S(\textbf{Q})\equiv (S(\textbf{Q})_p-S(\textbf{Q})_0)/S(\textbf{Q})_0,$ is shown in the left half of Fig.~\ref{fig:dSQ001}, simulated at $T=1.5\textup{ K}$ and $p=1.5\textup{ GPa}$. In the right half we show the change in the experimental\cite{dataThatWeTookOnILL_The001Direction} scattering at $p=2.2\textup{ GPa}$ and $T=1.5\textup{ K}$ measured in previous work\cite{OurArticle}. In the outer region we see eight patches of increased intensity which have been marked and coincide with the theory prediction. In the inner region there are two regions of increased intensity not seen in the the experiment, but we note, as mentioned in previous work, that the experimentally measured signal was poorly sampled at wave vectors shorter than $1.5 \textup{ \AA\textsuperscript{-1}}$ due to windows in the scattering plane of the previous CuBe pressure cell\cite{OurArticle} used in that neutron scattering experiment.

\begin{figure}[h!]
    \centering
    \includegraphics[width=0.85\linewidth]{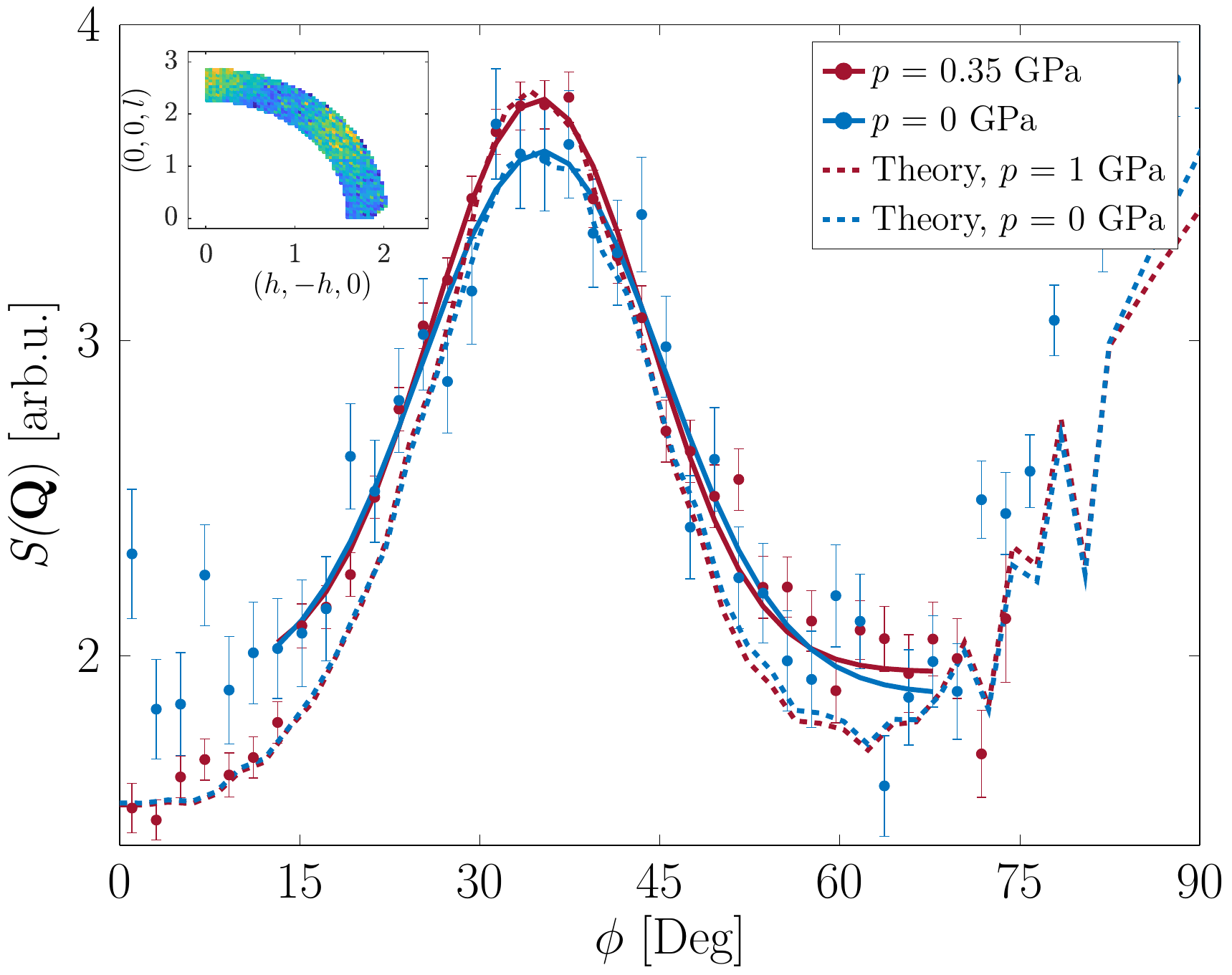}
    \caption{$[110]$ spin-flip $S(\textbf{Q})$ line-cut average over $2.22< Q < 2.87$ (reduced units), for polar angle $\phi= 0^\circ\rightarrow 90^\circ$ in the neutron scattering data (shaded area in Fig.\ref{fig:NeutronScatteringExperiment110})\cite{dataThatWeTookOnILL_The110Direction}. Gaussian fit to the experimental data (solid). Theoretical prediction at zero and $1.0 \textup{ GPa}$ uniaxial pressure for the parameters found in the measurements of the magnetic moment, $J_\parallel=1.2\textup{ K}, \kappa=1\%$, Fig.~\ref{fig:delta001.111.110}(c) (dashed). MC simulation for $8192$ particles $L=8$.}
    \label{fig:110NeutronSlice}
\end{figure}

A key result of this study is therefore that the scattering profile calculated from the parameters obtained from the fit to the magnetic moment, Fig.~\ref{fig:delta001.111.110}(a), matches the best profile that can be obtained by freely adjusting $J_\parallel$ to the experimental neutron data, as was done in the previous study\cite{OurArticle}. Furthermore, we use the DSPM to calculate the spin-flip structure factor in the $(h,-h,l)$ plane with pressure along $[110]$, shown in Fig.~\ref{fig:NeutronScatteringExperiment110}. We take the parameters fitted to the magnetic moment, Fig.~\ref{fig:delta001.111.110}(c), and calculate $S(\textbf{Q})$ from Eq.~(\ref{Eq.SFstructureFactor}). In order to better show the changes in scattering intensity, we integrate the signal around one of the satellite peaks as shown by the shaded region in Fig.~\ref{fig:NeutronScatteringExperiment110}. In Fig.~\ref{fig:110NeutronSlice} we see Gaussian fits to the experimental\cite{dataThatWeTookOnILL_The110Direction} data (solid lines). The theoretical estimate from the fit to the magnetic moment ($J_\parallel=1.2\textup{ K}$,$\kappa=1\%$) is shown together with the theoretical prediction at zero pressure (dashed lines). Although the experimental error bars, set by the neutron exposure time, are of the same order as the observed increase under applied pressure, we note that the theory captures the increased peak intensity of about 4 $\%$ and conclude that the model describes the observed phenomena both for magnetization and neutron scattering measurements in several crystalline directions.

\subsection{Evolution of the model parameters}\label{Section:EvolutionofParameters}

\begin{figure}[!h]
	\centering
	\includegraphics[width=0.9\linewidth]{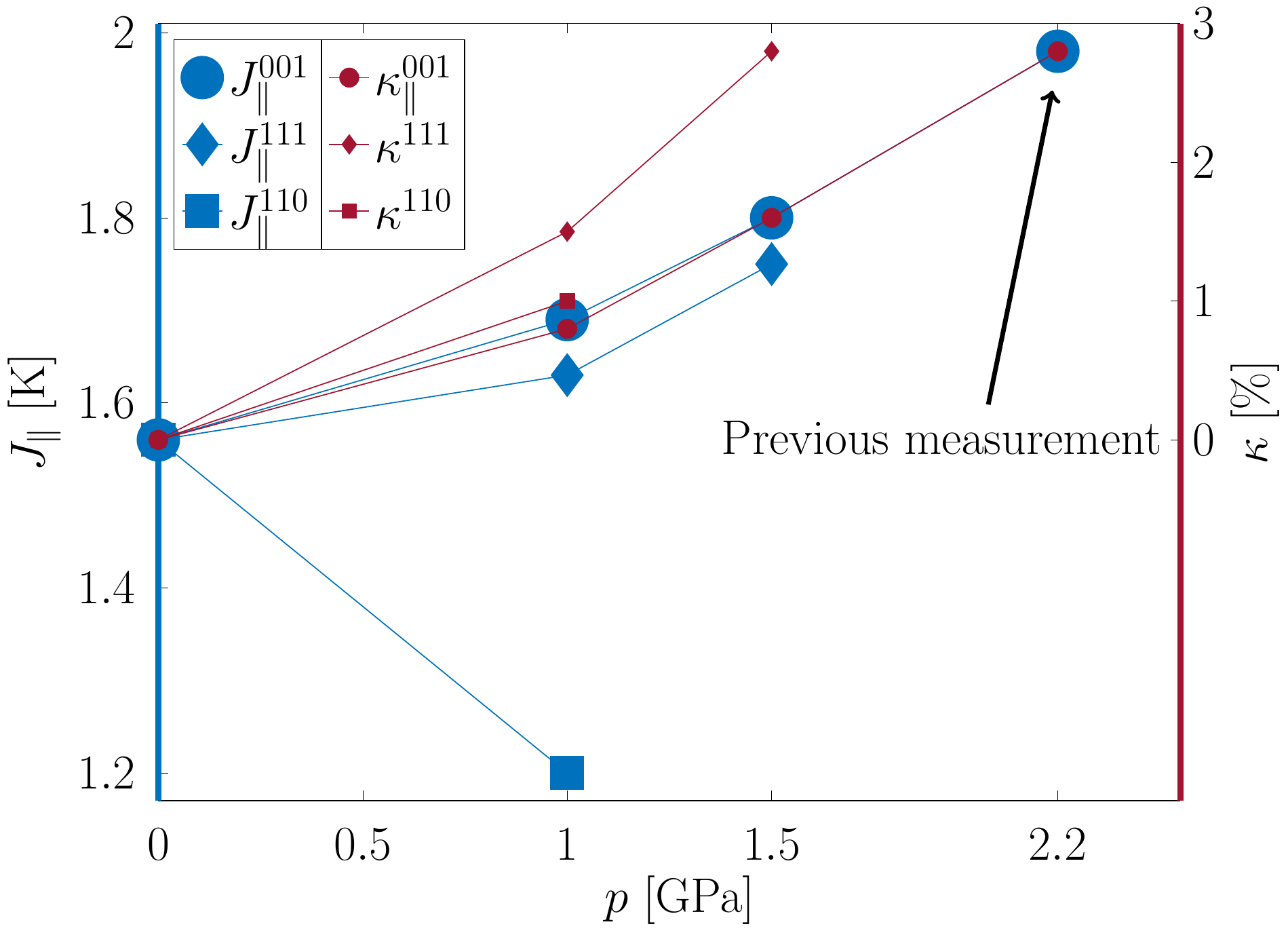}
	\caption{Evolution of the model parameter $J_\parallel$ and crystal compression $\kappa$ under applied uniaxial pressure along the $[001]$ crystalline axis.}
	\label{EvolutionOfJ}
\end{figure}

Using the straight forward magnetization measurements we can obtain the model parameter dependence on pressure. In Fig.~\ref{EvolutionOfJ} we depict the pressure dependence of $J_\parallel$ and $\kappa$ in all three direction. We include the current data points at pressures of $1.0\textup{ GPa}$ and $p=1.5\textup{ GPa}$, as well as the result of the previous neutron study at an applied pressure of $2.2\textup{ GPa}$\cite{OurArticle}. The three data points along $[001]$ show a near-linear dependence on pressure. Analysing the phase diagram using a combination of MC and direct comparison of the state energies\cite {OurArticle}  we find that all three points lie on a curve in the $(J_\parallel,\kappa)$-space for which the exchange interaction evolves as to cancel the changes in the dipolar dynamics originating from lattice compression, and  the system is on the border between two different types of dipolar chain\cite{MelkoGingras} ground states\cite{OurArticle}. This suggests that the ground state will remain a dipolar chain state under application of uniaxial pressure along the $[001]$ direction. This result contrasts to the predictions for DTO, where at sufficiently high pressure a ferromagnetic ground state is expected\cite{Jaubert1,OurArticle}.

In order to extrapolate our results to higher pressure we use  linear extrapolations for the parameters in the  $[111]$ and $[110]$ directions:
\begin{equation}\label{eq:linearExtrapolation}
\begin{split}
      J_\parallel&=J_\parallel^{p=0}+(J_\parallel^{p^*}-J_\parallel^{p=0})p,\\
      \kappa_\parallel&=\kappa^{p=0}+(\kappa^{p^*}-\kappa^{p=0})p.
\end{split}
\end{equation}
For $\kappa$ this is  motivated by Hooks law, while the true pressure dependence on  $J_\parallel$  may deviate from our linear model, but we proceed with $p^*$ as the highest measured pressure for the respective direction.

For the $[111]$ direction, we find that the dipolar chain ground state is stable up to pressures of at least $4\textup{ GPa}$. In the $[110]$ direction, on the other hand,  we find a transition to the previously mentioned ferromagnetic state  at a critical pressure of  $3.3\textup{ GPa}$. In DTO this state is expected when applying pressure in excess of $3.4\textup{ GPa}$ in the $[001]$ direction\cite{Jaubert1,OurArticle}  We therefore conclude that further uniaxial high pressure studies should be conducted either on DTO in the $[001]$ direction, or on HTO in the $[110]$ direction. At such high pressures the crystals are likely to break. However, using techniques like submerging the crystals in epoxy  resin as we did in this study, or using some other type of support material, we believe that such experiments can be realized. We summarize the fitted and estimated parameter values in table~\ref{table:parametersSummaryTable}.


\begin{table}[!h]
\begin{tabular}{|l|l|ll|ll|l|}
\hline
\multirow{2}{*}{Direction} & $p=0\textup{ GPa}$ & \multicolumn{2}{l|}{$p=1\textup{ GPa}$} & \multicolumn{2}{l|}{$p=1.5\textup{GPa}$} & \multirow{2}{*}{$p_c\textup{ [GPa]}$} \\ \cline{2-6}
                           & $J\textup{ [K]}$      & $J_\parallel\textup{ [K]}$         & $\kappa\textup{ [\%]}$        & $J_\parallel\textup{ [K]}$          & $\kappa\textup{ [\%]}$         &                     \\ \hline
{[}001{]}                  & 1.56   & 1.69         & 0.8          & 1.80          & 1.6           & -                   \\
{[}111{]}                  & 1.56   & 1.63         & 1.5          & 1.75          & 2.8           & $>4$ \\
{[}110{]}                  & 1.56   & 1.2          & 1            & -             & -             & $3.3$             \\ \hline
\end{tabular}
	\caption{Summary of the fitted and estimated DSPM parameters found in this work. $p_c$ is the critical pressure above which a linear extrapolation predicts a ferromagnetic\cite{Jaubert1} ground state. We estimate the error to be less than $5\%$ of the fitted parameter value. However, for the [110] direction this might be as much as $20\%$ due to misalignment having more prominent effects.}\label{table:parametersSummaryTable}
\end{table}

\section{Conclusions}

We have performed measurements of the field-induced magnetic moment, magnetic susceptibility and neutron structure factor of HTO under applied uniaxial pressure. Through extensive MC calculations we demonstrate that a dipolar spin ice model, with a pressure-tuned nearest neighbor interaction is able to capture the most essential features of the measurements of HTO. The framework is  extended  to include effects of misalignment, and we have found that misalignment can, to some extent, describe the anomalous effects observed in the magnetic moment for the $[110]$ direction in both HTO and DTO. 

The $T$-dependent pressure induced changes in the susceptibility turns out to be a more sensitive quantity to both measure and model than the field-induced magnetic moment. The low field used experimentally results in a weak signal which is easily overshadowed. Theoretically, the susceptibility of spin ice has proven to be a sensitive function of the intrinsic competing interactions\cite{bovo18}, and the measurements are sensitive to sample shape\cite{tweng17}, with the optimal sample size probably spherical. To perform susceptibility measurements on a spherical sample under pressure is highly challenging, and we therefore have to contend with our present results. Still, the qualitative curve shape for the susceptibility in the $[001]$ direction is captured rather well by our model, but there are significant discrepancies in the $[111]$ measurement.

Our main result is that the model parameters derived from the measurement of the magnetization also captures the most salient features of the pressure induced change in the neutron scattering structure factor. That the same model describes both bulk properties and spin-spin correlation functions lends credibility to the theory. Therefore we hope that using relatively straight forward magnetization measurements to determine the pressure dependence of interaction parameters can prove useful also when it comes to other classes of frustrated materials. Increased theoretical predictive power supporting demanding neutron experiments under high pressure would benefit many investigators in the field. With new, and more intense, neutron sources under construction we expect this to be a research topic of increasing importance in the near future.

\appendix
\section{Misalignment}\label{Appendix:misalignment}
\begin{figure}[h!]
    \centering
    \includegraphics[width=0.4\textwidth]{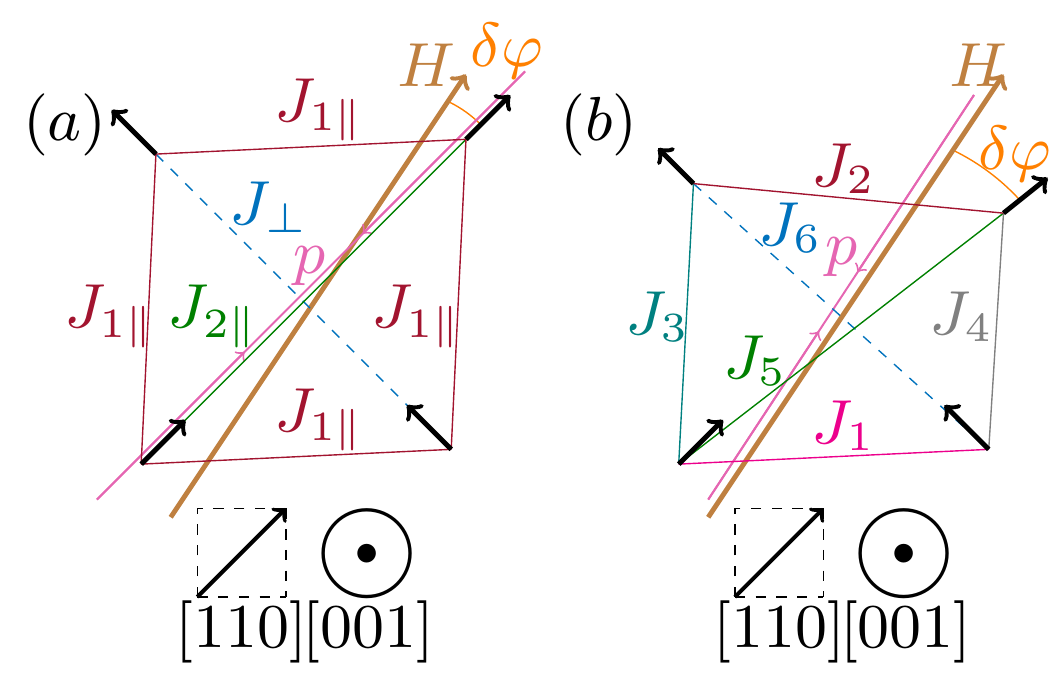}
    \caption{Misalignment. (a) Field misalignment, discussed in main text. (b) Crystalline axis misalignment. The view is along the [001] direction in both cases. More symmetries are broken in (b) as the tetrahedron is compressed along the $p$ axis.}
    \label{fig:extendedMisalignment}
\end{figure}
The main text discusses the influence of misalignment of the applied field. Of even greater importance is probably a misalignment of the crystalline axis with respect to the applied pressure and field.  Our basic analysis for the magnetization measurements suggests that the $[001]$ and $[111]$ directions are not sensitive to a slight misalignment. The $[110]$ direction, on the other hand, is much more sensitive due to the large subset of spins that are perpendicular to the field under perfect alignment. We suspect that this result holds also in more elaborate models of crystalline axis misalignment. A more general description than provided in the main text introduces up to six independent exchange parameters since there are six different types of nearest neighbor bonds in a tetrahedron. Figure~\ref{fig:extendedMisalignment} illustrates the two different kinds of misalignment. The field misalignment depicted in  Fig.~\ref{fig:extendedMisalignment}(a), was discussed in the main text. In this case, there are only three distinct exchange couplings ($J_\perp$,$J_{1\parallel}$,$J_{2\parallel}$) due to  the symmetry of the tetrahedron. In the case of misalignment of the  the crystalline axis the $J_{2\parallel}$ couplings will no longer be equal. Figure~\ref{fig:extendedMisalignment}(b) illustrates a case where all six distances between different corners of the compressed tetrahedron are different. In order to reduce the number of free parameters, we could linearize the the distance dependence of $J$, $J=J_{p=0}+K(r_0-r)$, where $r_0$ is the unperturbed distance and $K$ is a free parameter. This would give the same number of free parameters as in the basic model presented in the main text. However, $J$ does not necessarily depend on the distance between the ions in the same way for all nearest neighbor bonds. Therefore, we contend with the model of the main text, and believe that it sufficiently demonstrates that the features above $1\textup{ T}$ in the $[110]$ relative change in magnetic moment can be an artefact due to misalignment.

\section{Crystals}

\begin{figure}[h!]
\begin{subfigure}[b]{.22\textwidth}
    \centering
    \includegraphics[width=0.8\textwidth]{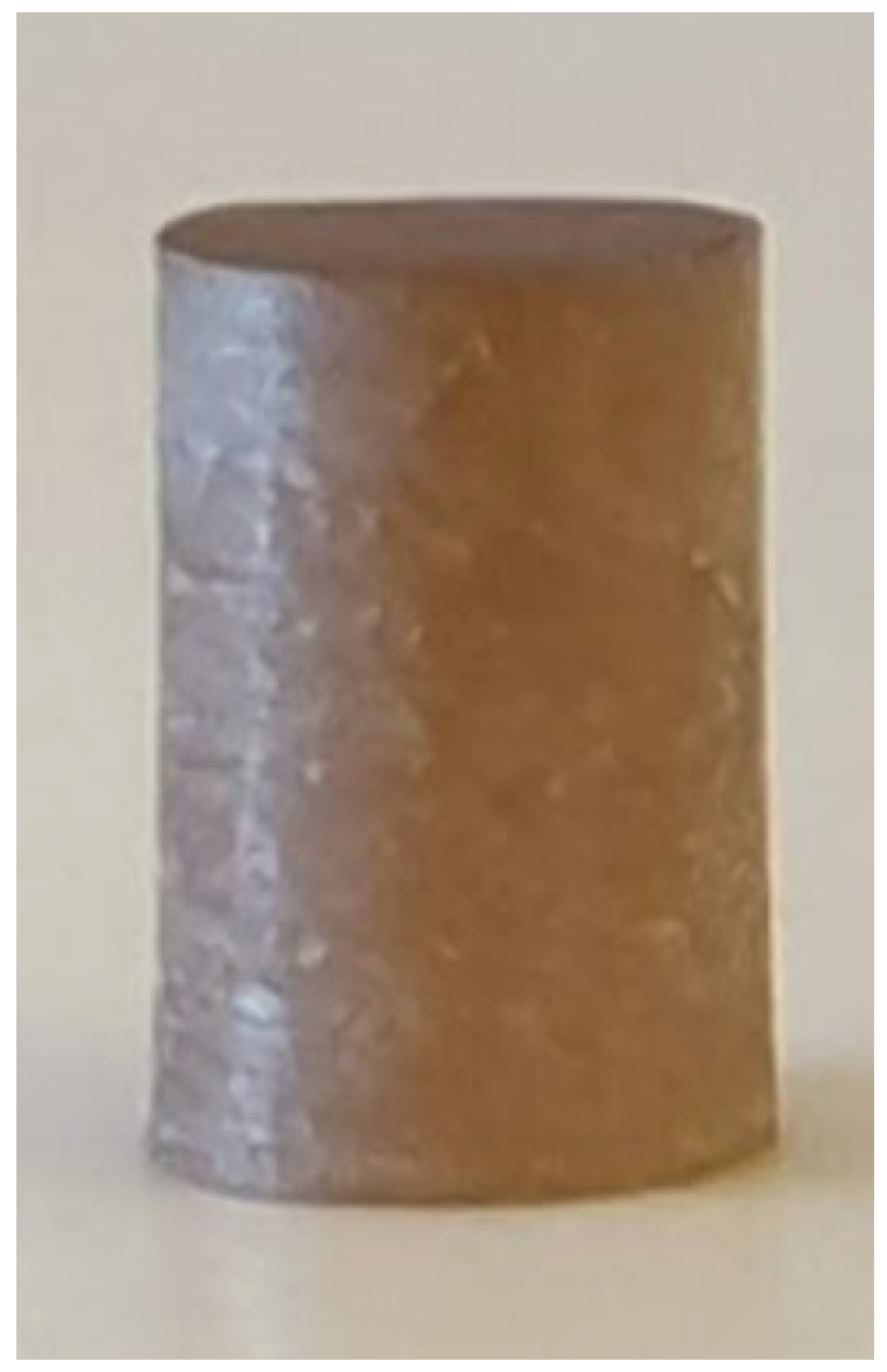}
    \caption{}\label{fig:Cupid001MisalignmentPicutre}
\end{subfigure}
\begin{subfigure}[b]{.22\textwidth}
    \centering
    \includegraphics[width=1.09\textwidth]{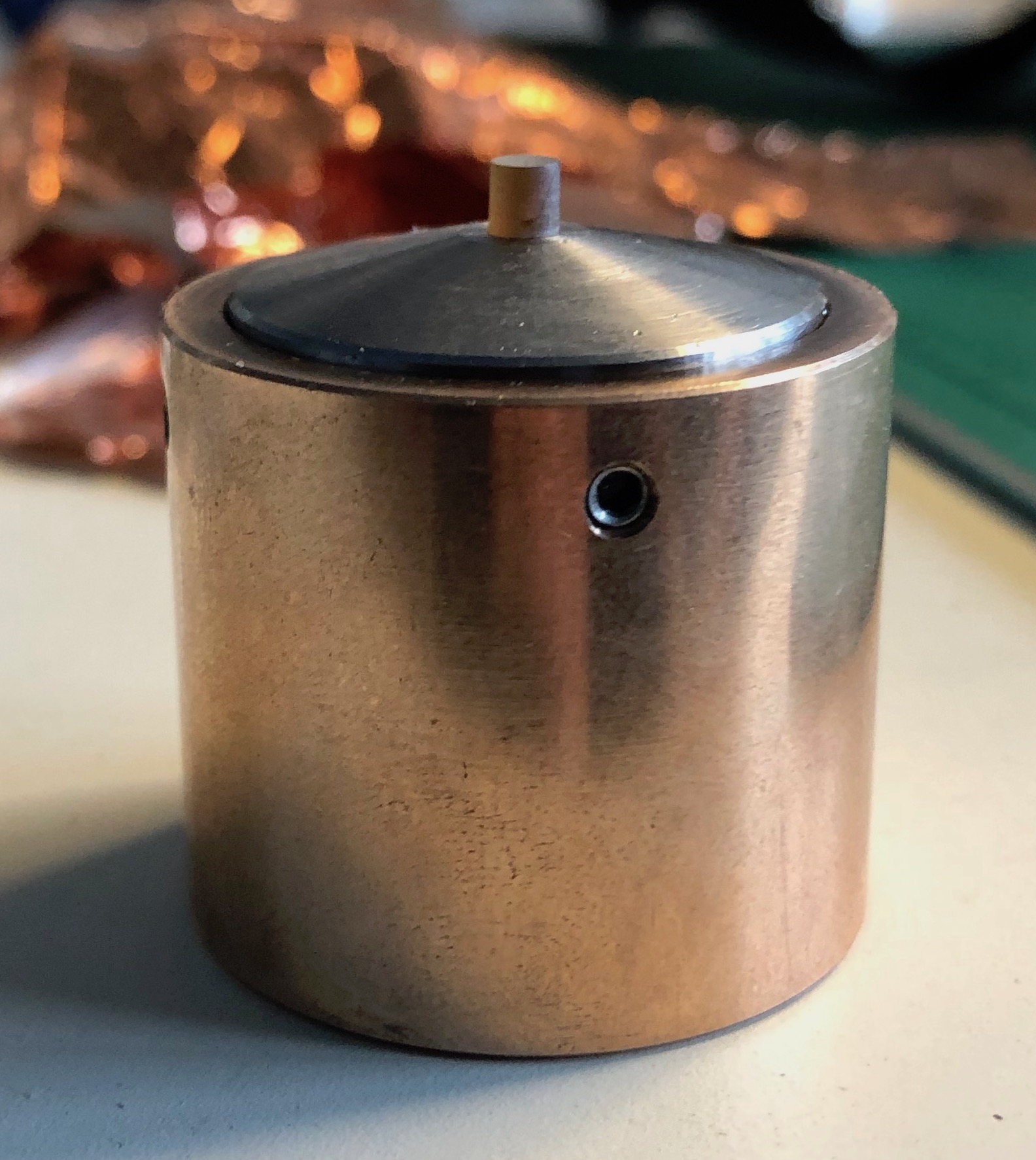}
    \caption{}\label{fig:NeutronCrystal}
\end{subfigure}
\caption{(\subref{fig:Cupid001MisalignmentPicutre}) One of the HTO crystals ($\diameter~2\textup{ mm}$) used for the magnetization measurement. The picture shows the accuracy of the cutting procedure. (\subref{fig:NeutronCrystal}) One of the crystals used in the neutron scattering experiment ($\diameter~3\textup{ mm}$), placed on top of the steel anvil which was used for applying uniaxial pressure.}
\end{figure}

The crystal quality is particularly important when working with external pressure. The crystals must withstand high pressure without cracking, and due to the demagnetizing effects it is important that the crystals are cut with accuracy. Figure~\ref{fig:Cupid001MisalignmentPicutre} shows a picture of one of the crystals used in the magnetization experiment. The crystals have been aligned to the uniaxial pressure axis. However as the crystals were submerged in stycast which is much softer than HTO, a small shift in the alignment may arise as pressure is applied due to plastic deformation. 

In the neutron scattering experiment for the larger $3\textup{ mm}$ diameter crystals, the nuclear Bragg peaks observed on D7 are consistent with the alignment. However, the instrumental parameters of D7: incident beam size, divergence and detection resolution, do not provide a very accurate determination of the absolute alignment. The nature of the neutron scattering experiment also prevents the use of stycast to mitigate against the Poisson expansion. To include such expansion of the sample in the simulation would introduces additional unknown model parameters, which from these few measurements is hard to determine. We therefore chose not to include this effect, and we find that this approximation still gives valid prediction for the neutron scattering intensity. However, we mention that the physics that can be reached by including this effect can also be reached by allowing $J_\perp$ to vary, and it will not introduce any new effects, apart from having more free parameters allowing for an easier fit to data.


\section{Uniaxial pressure cell}\label{Appendix:pressurecell}
\begin{figure}[h!]
    \centering
    \includegraphics[width=0.4\textwidth]{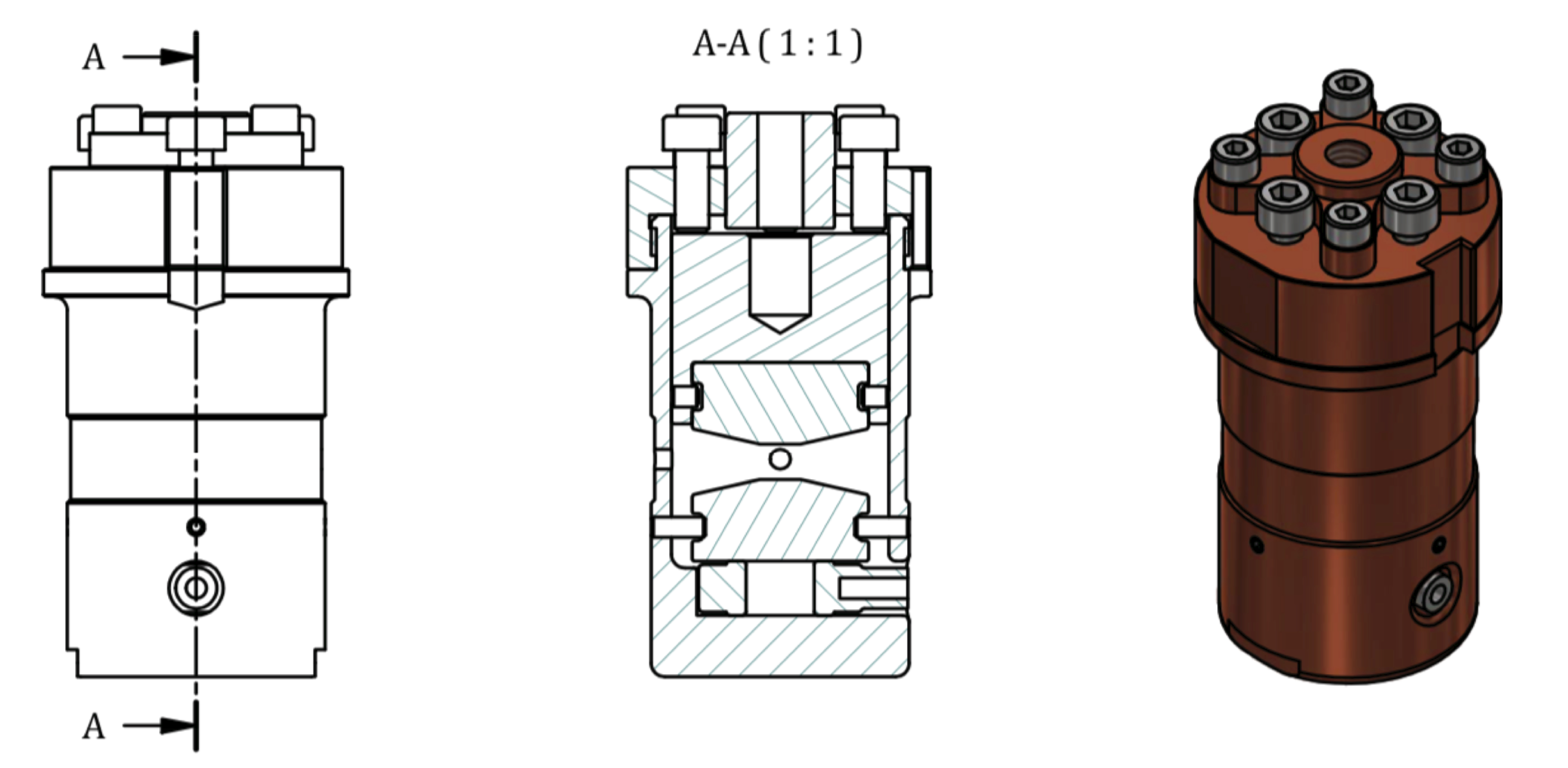}
    \caption{Engineering drawings and overview of the uniaxial pressure cell.}
    \label{fig:PressureCellB}
\end{figure}

The neutron scattering uniaxial pressure cell has been developed to enable neutron scattering experiments in the cold energy spectra range, $2\lesssim\lambda~\lesssim20~\textup{\AA}$, for diffuse magnetic scattering profiles, inclusive of weak inelastic scattering features, with polarisation analysis. The uniaxial pressure cell should be able to provide pressures up to $2\textup{ GPa}$ at cryogenic temperatures. As such, the requirements include a scattering window that covers a wide angular range with a very clean background profile and a non-magnetic cell that enables polarisation analysis. These requirements place stringent restrictions on the materials for the manufacturing of the cell and lead us to focus on the optimization of an anvil type cell. An engineering overview of the cell employed during the D7 experiment is shown in Fig.~\ref{fig:PressureCellB}. The main body is manufactured out of CuBe with non-magnetic stainless steel anvils. Force is applied at room temperature and recalibrated using a calibration profile developed within our team. Pressure is deduced from the known contact area. The calibration profiles are not perfect and may lead to some uncertainty in the exact pressure applied at the lowest temperatures. Complete details of the cell will be published elsewhere and an  uniaxial pressure cell with in situ pressure determination is under development ensuring we maintain the aforementioned parameters. 

\acknowledgements
We thank Steven Bramwell for insightful discussions. 
The neutron scattering experiments were performed at the Paul Scherrer Institute and the Institute Laue-Langevin. We thank the sample environment group of the ILL for preparatory help and access to facilities. The simulations were performed on resources provided by the Swedish National Infrastructure for Computing (SNIC) at the Center for High Performance Computing (PDC) at the Royal Institute of Technology (KTH). The project was supported by Nordforsk through the program NNSP (Project No. 82248) and by the Danish Agency for Research and Innovation through DANSCATT.

\end{document}